\documentclass[12pt,draftclsnofoot, onecolumn]{IEEEtran}

\usepackage{acronym}
\usepackage{tikz}
\usepackage{times}
\usepackage{amsmath,dsfont}
\usepackage{amssymb,amsthm}
\usepackage{epsfig,verbatim}
\usepackage{subfigure}
\usepackage{setspace}
\usepackage{color}
\usepackage{epstopdf}
\usepackage{graphics}
\usepackage{accents}
\usepackage{booktabs}
\usepackage{mathtools}
\usepackage{algorithm}
\usepackage{algorithmic}
\usepackage{enumitem}
\usetikzlibrary{matrix,decorations.pathreplacing}

\newcommand{\myVec}[1]{{\boldsymbol{#1}}}
\newcommand{\myMat}[1]{{\boldsymbol{#1}}}
\newcommand{\mySet}[1]{\mathcal{#1}}
\newcommand{\myI}{{\myMat{I}}}			 		
	
\newcommand{\myY}{{\myVec{Y}}}			 		
\newcommand{\myS}{{\myVec{S}}}			 		

\newcommand{\Pdf}[1]{p_{ { #1}} }
\newcommand{\PdfEst}[1]{\hat{p}_{ { #1}} }
\newcommand{\Mem}{l}			 			
\newcommand{\Blklen}{t}			 			
\newcommand{\Blkset}{\mySet{T}}
\newcommand{\CnstSize}{m}			 			

\newcommand{\CovMat}[1]{\myMat{\Sigma}_{#1}}			

\newcommand{\Vecdim}[1]{} 

\newcommand{\eqspace}{\vspace{-0.2cm}}

\newcommand{\Nusers}{K}
\newcommand{\Nantennas}{n_r}
\newcommand{\NusersSet}{\mySet{K}}

\newcommand{\SigW}{\sigma_w^2}
\newcommand{\SigE}{\sigma_e^2}
\newcommand{\Niter}{Q}
\newcommand{\NiterSet}{\mySet{Q}}
\newcommand{\Ntraining}{n_t}

\theoremstyle{plain}

\theoremstyle{definition}

\theoremstyle{definition}

\newcommand{\figSpace}{\vspace{-0.6cm}}
\newcommand{\includefig}[1]{\includegraphics[width = 0.6\columnwidth]{#1} 	\vspace{-0.2cm}}
\acrodef{adc}[ADC]{analog-to-digital convertor}
\acrodef{cs}[CS]{compressed sensing}
\acrodef{dtft}[DTFT]{discrete-time Fourier transform}
\acrodef{dnn}[DNN]{deep neural network} 
\acrodef{csi}[CSI]{channel state information}
\acrodef{map}[MAP]{maximum a-posteriori probability}
\acrodef{snr}[SNR]{signal-to-noise ratio}
\acrodef{bs}[BS]{base station} 
\acrodef{iot}[IOT]{Interent of Things}
\acrodef{mimo}[MIMO]{multiple-input multiple-output}
\acrodef{mse}[MSE]{mean-squared error}
\acrodef{mmse}[MMSE]{minimum \ac{mse}}
\acrodef{pdf}[PDF]{probability density function}
\acrodef{rv}[RV]{random variable}
\acrodef{fec}[FEC]{forward error correction}
\acrodef{rs}[RS]{Reed-Solomon}
\acrodef{lti}[LTI]{linear time-invariant}
\acrodef{wss}[WSS]{wide-sense stationary}
\acrodef{psd}[PSD]{power spectral density}
\acrodef{ser}[SER]{symbol error rate} 
\acrodef{ber}[BER]{bit error rate} 
\acrodef{sgd}[SGD]{stochastic gradient descent} 
\acrodef{isi}[ISI]{intersymbol interference}  
\acrodef{awgn}[AWGN]{additive white Gaussian noise} 
\acrodef{ut}[UT]{user terminal} 
\acrodef{mmw}[mmWave]{millimeter wave}
\acrodef{cfl}[CFL]{clustered \ac{fl}} 
\acrodef{fl}[FL]{federated learning}
\acrodef{pic}[PIC]{principal inertia component}
\acrodef{ml}[ML]{machine learning}
\acrodef{lstm}[LSTM]{long short-term memory} 
\acrodef{em}[EM]{expectation minimization} 
\acrodef{awgn}[AWGN]{additive white Gaussian noise}
\acrodef{rnn}[RNN]{recurrent neural network}
\acrodef{sbrnn}[SBRNN]{sliding bidirectional \ac{rnn}}
\acrodef{sic}[SIC]{soft interference cancellation}
\acrodef{hmm}[HMM]{hidden Markov model}
\acrodef{bpsk}[BPSK]{binary phase shift keying}

\newcommand{\revision}[1]{#1}

\IEEEoverridecommandlockouts 

\title{Data-Driven Symbol Detection via Model-Based Machine Learning}

\author{
	\IEEEauthorblockN{Nariman Farsad, Nir Shlezinger, Andrea J. Goldsmith and Yonina C. Eldar 
	}  
	\thanks{This work was supported in part by the  US - Israel Binational Science Foundation	under grant No. 2026094,  by the Israel Science Foundation under grant No. 0100101, and by the Office of the Naval Research under grant No. 18-1-2191.
	}
	\thanks{ 
	N. Farsad  and A. J. Goldsmith are with the Department of EE, Stanford, Palo Alto, CA (e-mail:  nfarsad@stanford.edu; andrea@wsl.stanford.edu).  	
}	
	\thanks{
		N. Shlezinger  and Y. C. Eldar are with the Faculty of Math and CS, Weizmann Institute of Science, Rehovot, Israel (e-mail: nirshlezinger1@gmail.com; yonina@weizmann.ac.il). 	
	}

	\vspace{-1.0cm}
}

\begin{document}

 	\maketitle
 \pagestyle{plain}
 \thispagestyle{plain}

\begin{abstract}
	The design of symbol detectors in digital communication systems has traditionally relied on statistical channel models that describe the relation between the transmitted symbols and the observed signal at the receiver. Here we review a data-driven framework to symbol detection design which combines machine learning (ML) and model-based algorithms. In this hybrid approach, well-known channel-model-based algorithms such as the Viterbi method, BCJR detection, and multiple-input  multiple-output (MIMO) soft interference cancellation (SIC) are augmented with ML-based algorithms to remove their channel-model-dependence, allowing the receiver to learn to implement these algorithms solely from data. The resulting data-driven  receivers are most suitable for systems where the underlying channel models are  poorly understood, highly complex, or do not well-capture the underlying physics. Our approach is unique in that it only replaces the channel-model-based computations with dedicated neural networks that can be trained from a small amount of data, while keeping the general algorithm intact. Our results demonstrate that these techniques can yield near-optimal performance of model-based algorithms without knowing the exact channel input-output statistical relationship and in the presence of channel state information uncertainty.  
\end{abstract}
 
\vspace{-0.2cm}
\section{Introduction}
\vspace{-0.2cm}
	In digital communication systems, the receiver is required to reliably recover the transmitted symbols from the observed channel output. This task is commonly referred to as {\em symbol detection}. Conventional symbol detection algorithms, such as those based on the \ac{map} rule, require complete knowledge of the underlying channel model and its parameters \cite{goldsmith2005wireless,tse2005fundamentals}. 
Consequently, in some cases, conventional methods cannot be applied, particularly when the channel model is highly complex, poorly understood, or does not well-capture the underlying physics of the system. 
Furthermore, when the channel models are known, many detection algorithms rely on \ac{csi}, i.e., the instantaneous parameters of the channel model, for detection. Therefore, conventional channel-model-based techniques require the instantaneous \ac{csi} to be estimated. This  process entails overhead, which decreases the data rate. Moreover, inaccurate  \ac{csi} estimation typically degrades the detection performance.

An alternative data-driven approach to model-based algorithms uses \ac{ml}. \ac{ml} methods, and in particular, \acp{dnn}, have been the focus of extensive research in recent years due to their empirical success in various applications, including image processing and speech processing \cite{lecun2015deep}. 
\ac{ml} schemes have several benefits over traditional model-based approaches: First, \ac{ml} methods are independent of the underlying stochastic model, and  can operate efficiently in scenarios where this model is unknown or its parameters cannot be accurately estimated.
Second, when the underlying model is extremely complex, \ac{ml} algorithms have demonstrated the ability to extract  meaningful semantic information from the observed data \cite{bengio2009learning}, a task which is very difficult to carry out using  model-based methods. 
Finally, \ac{ml} techniques often lead to faster inference compared to iterative model-based approaches, even when the model is known \cite{gregor2010learning, monga2019algorithm}.

Recent years have witnessed growing interest in the application of \acp{dnn} for receiver design;  see detailed surveys in \cite{oshea2017introduction, simeone2018very, mao2018deep, gunduz2019machine, balatsoukas2019deep}. 
Unlike model-based receivers, which implement a specified detection rule, \ac{ml}-based receivers learn how to map the channel outputs into the transmitted symbols from training, namely, they operate in a data-driven manner. 
Broadly speaking, previously proposed \ac{ml}-based receivers can be divided into two main categories: Conventional \acp{dnn} and unfolded networks \cite{monga2019algorithm}.  The first group replaces the receiver processing with a \ac{dnn} architecture from established methods in the \ac{ml} literature. A-priori knowledge of the channel model is accounted for in the selection of the network type, which is typically treated as a black box. For example, \acp{rnn} were applied for  decoding sequential codes in \cite{kim2018communication}; the work \cite{farsad2018neural} used sliding bi-directional \acp{rnn} for \ac{isi} channels;  and the work \cite{caciularu2018blind} used variational autoencoders for unsupervised equalization. Such \acp{dnn},  which use conventional network architectures that are ignorant of the underlying channel model, can typically operate reliably in various scenarios with or without \ac{csi} and channel model knowledge, assuming that they were properly trained for the specific setup. Nonetheless, block box \acp{dnn} tend to have a large number of parameters, and thus require large data sets to train \cite{balatsoukas2019deep}, limiting their application in dynamic environments, which are commonly encountered in communications. 

Unlike conventional \acp{dnn}, which utilize established architectures, in unfolded receivers the network structure is designed following a model-based algorithm. In particular, deep unfolding is a method for converting an iterative algorithm into a \ac{dnn} by designing each layer of the network to resemble a single iteration \cite{hershey2014deep,monga2019algorithm,solomon2019deep}. The resulting \ac{dnn} tends to demonstrate improved convergence speed and robustness compared to the model-based algorithm \cite{gregor2010learning, monga2019algorithm}. In the context of   symbol detection, the works \cite{samuel2019learning, takabe2019deep} designed deep receivers by unfolding the projected gradient descent algorithm for recovering the \ac{map} solution, while \cite{he2018model} unfolded the approximate message passing optimization scheme, and \cite{khobahi2019deep} proposed to recover signals obtained from one-bit quantized measurements by unfolding gradient descent optimization. Compared to conventional \acp{dnn}, unfolded networks are typically interpretable, and tend to have a smaller number of parameters, and can thus be trained quicker \cite{balatsoukas2019deep}. However,  these previously proposed receivers all assume linear channel models with Gaussian noise, in which \ac{csi} is available \cite{samuel2019learning, takabe2019deep, he2018model,khobahi2019deep}. Consequently, these methods do not capture the potential of \ac{ml} in being independent of the model, and thus are applicable only under specific  setups.  

In this work, we design and study \ac{ml}-based symbol detection, focusing and presenting some of our recent work in this area in a tutorial fashion \cite{shlezinger2019viterbinet, shlezinger2020data, shlezinger2019deepSIC}. The main distinction between our approach and prior work is using \ac{ml} in conjunction with well-known detection algorithms in a hybrid fashion. Specifically, our design is based on integrating \ac{ml} methods into established detection algorithms, allowing them to be learned from training and operate in a data-driven fashion. 
Our approach to symbol detection implements well-known channel-model-based algorithms such as the Viterbi algorithm \cite{viterbi1967error}, the BCJR method \cite{bahl1974optimal}, and \ac{mimo} \ac{sic} \cite{choi2000iterative}, while only removing their channel-model-dependence by replacing the \ac{csi}-based computations with dedicated \acp{dnn}, i.e., we {\em integrate \ac{ml} into these channel-model-based algorithms to to learn CSI}. 

We begin by presenting in detail our approach for combining \ac{ml} and model-based algorithms, and its relationship with deep unfolding. Then we  present two channel models to which we apply our approach for symbol detection:
First, we focus on finite-memory channels and present ViterbiNet \cite{shlezinger2019viterbinet} and  BCJRNet \cite{shlezinger2020data}, which are data-driven implementations of the maximum-likelihood sequence detector and the \ac{map} symbol detector for this channel, respectively. We then focus on the multiuser \ac{mimo} channel and present DeepSIC \cite{shlezinger2019deepSIC}, which learns to implement the iterative \ac{sic} algorithm under this channel model. Our resulting receivers are shown to approach the performance of their model-based counterparts without requiring \ac{csi} and using relatively small training sets. We also observe that, in the presence of \ac{csi} uncertainty, the performance of the model-based algorithms is significantly degraded, while the combined \ac{ml} and model-based approach is capable of reliably detecting the symbols.  Our results demonstrate the potential gains of  combining \ac{ml} and model-based algorithms over both pure model-based and pure ML-based methods.


The rest of this paper is organized as follows: In Section~\ref{sec:ModelML} we describe the general hybrid \ac{ml} approach. Section~\ref{sec:FiniteMemChan} focuses on finite-memory channels and presents the ViterbiNet and the BCJRNet for these channels.  Section~\ref{sec:DeepSIC} details the data-driven \ac{sic} algorithm, and Section~\ref{sec:Conclusion} provides concluding remarks.

\vspace{-0,2cm}
\section{Model-Based Machine Learning}
\label{sec:ModelML}
\vspace{-0,2cm}
In this section we present our proposed methodology for converting  model-based algorithms to data-driven techniques by integrating \ac{ml} into them. We focus here on the high-level rationale, while concrete examples for  symbol detection methods are presented in Sections \ref{sec:FiniteMemChan}-\ref{sec:DeepSIC}. In particular, we first motivate the fusion of \ac{ml} and channel-model-based symbol detection schemes in Section \ref{subsec:Rat_DatavsModel}. Then, in Section \ref{subsec:Rat_DeepInteg} we present how \ac{ml} methods can be combined with model-based algorithms.

\vspace{-0.2cm}
\subsection{Machine Learning for Symbol Detection}
\label{subsec:Rat_DatavsModel}
\begin{figure}
	\centering
	\includegraphics[width = 0.8\columnwidth]{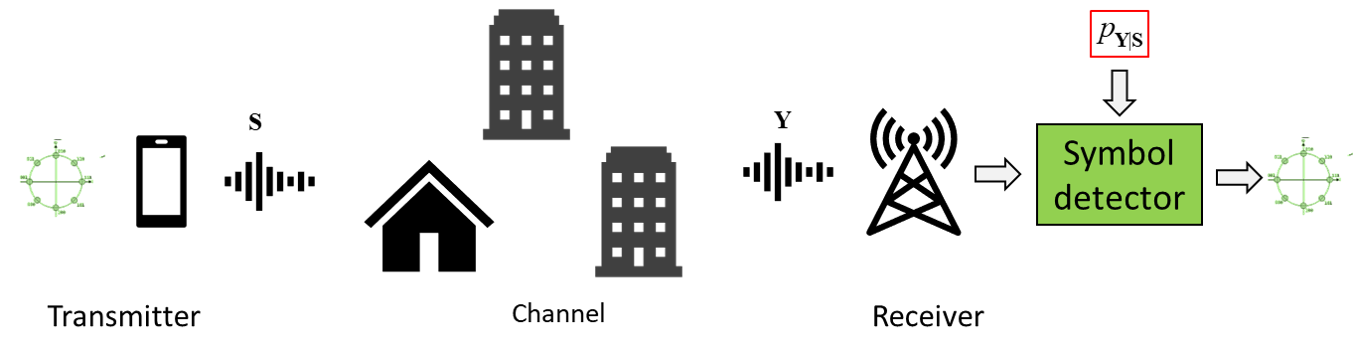} 
	\vspace{-0.6cm}
	\caption{Channel-model-based symbol detection illustration.}
	\label{fig:SymbolDetection1}
\end{figure}

\vspace{-0.2cm}
Symbol detection is arguably the most basic problem in digital communications. Symbol detection refers to how the receiver recovers a set of transmitted symbols denoted by $\myVec{S}$, which belong to some discrete constellation, from its observed channel output denoted by $\myVec{Y}$. This problem has been widely studied, and various algorithms with optimality guarantees and controllable complexity measures have been proposed over decades of research, depending on the statistical model relating $\myVec{Y}$ and  $\myVec{S}$, i.e., their conditional distribution $\Pdf{\myVec{Y} | \myVec{S}}$ \cite[Ch. 5]{goldsmith2005wireless}. The vast majority of these previously proposed schemes are {\em channel-model-based}, namely, they require accurate prior knowledge of $\Pdf{\myVec{Y} | \myVec{S}}$, and typically assume it takes some simplified form, e.g., that of a linear channel with \ac{awgn}. An illustration of conventional channel-model-based symbol detection is depicted in Fig. \ref{fig:SymbolDetection1}. As discussed in the introduction, these requirements limit the application of conventional model-based schemes in complex environments, and may entail substantial overhead due to the need to constantly estimate \ac{csi} at the receiver. 

Recent years have witnessed a dramatically growing interest in \ac{ml} methods, and particularly, in \acp{dnn}. These data-driven trainable structures have demonstrated unprecedented empirical success in various applications, including computer vision and speech processing \cite{lecun2015deep}. \ac{ml}-driven techniques have several key benefits over traditional model-based approaches: First, \ac{ml} methods are independent of the underlying stochastic model, and thus can operate efficiently in scenarios where this model is unknown, or its parameters cannot be accurately estimated \cite[Ch. 1]{Goodfellow-et-al-2016}. Second, when the underlying model is extremely complex, \ac{ml} algorithms have demonstrated the ability to extract and disentangle the meaningful semantic information from the observed data \cite{bengio2009learning}, a task which is very difficult to carry out using traditional model-based approaches, even when the model is perfectly known. Specifically, \ac{ml} tools have empirically demonstrated the ability to disentangle the relevant semantic information under complex analytically intractable statistical relationships. Furthermore, \ac{ml} techniques tend to generalize well, achieving good performance when operating under a statistical model which is different than the one under which they were trained, assuming that this difference is not too large \cite{neyshabur2017exploring}.  Finally, \ac{ml} algorithms often lead to faster inference compared to iterative model-based approaches, even when the model is known \cite{gregor2010learning}.

Nonetheless, not every problem can and should be solved using \acp{dnn}. While networks with a large set of parameters can realize a broad set of functions, making them suitable for a wide range of tasks, a large data set reflecting the task under which the system must operate is required to train these networks. For example, symbol detection over wireless communication channels is typically dynamic in nature. This implies that \ac{dnn}-based symbol detectors must be either trained anew for each channel realization, a task which may be infeasible when large labeled sets are required for training, or alternatively, that the data-driven receiver should be trained in advance using samples corresponding to a large set of expected channel conditions. In the latter, typically the resulting deep symbol detector is likely to achieve reasonable but non-optimal performance when inferring using samples corresponding to each of channels condition.

Our goal is to allow model-based symbol detection schemes to be applied in scenarios for which, due to either a complex statistical model or lack of its knowledge, these methods cannot be applied directly. This is achieved by combining \ac{ml} and model-aware algorithms via model-based \ac{ml}, as discussed in the sequel.

\vspace{-0.2cm}
\subsection{Combining \ac{ml} and Model-Based Algorithms}
\label{subsec:Rat_DeepInteg}
\begin{figure}
	\centering
	\includefig{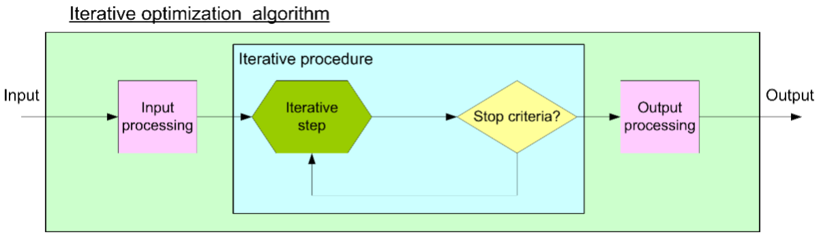} 
	\caption{Generic iterative algorithm illustration.}
	\label{fig:IterativeAlgo1}
\end{figure}

\vspace{-0.2cm} 
Here, we present our rationale for combining \ac{ml}, and in particular \ac{dnn}s, with channel-model-based symbol detection algorithms. 
Consider a symbol detection algorithm which produces an output vector $\hat{\myVec{S}}$ based on an observed input vector $\myVec{Y}$. Particularly, we focus on channels in which, given prior knowledge of $\Pdf{\myVec{Y} | \myVec{S}}$,  $\hat{\myVec{S}}$ can be recovered from $\myVec{Y}$ with provable performance and controllable complexity in an iterative fashion, e.g., via the Viterbi detector \cite{viterbi1967error} or the BCJR algorithm \cite{bahl1974optimal} for finite memory channels, or via interference cancellation methods for \ac{mimo} channels \cite{choi2000iterative}. This generic family of iterative algorithms  consists of some input and output processing stages, with an intermediate iterative procedure. The latter can in turn be divided into a channel-model-based computation, namely, a procedure that is determined by $\Pdf{\myVec{Y} | \myVec{S}}$; and a set of generic mathematical manipulations.
An illustration of this generic algorithmic procedure is depicted in   Fig. \ref{fig:IterativeAlgo1}.

Here, we want to implement the aforementioned algorithm in a data-driven manner. Possible motivations for doing so are that $p_{\myVec{S},\myVec{Y}}$ is unknown, costly to estimate accurately, or too complex to express analytically. A powerful tool for learning to implement tasks from data samples are \acp{dnn}, with their dramatic empirical success in applications such as computer vision, where the underlying model is very complex or poorly understood. 
Modern \ac{ml} suggests two main approaches for obtaining $\hat{\myVec{S}}$  from ${\myVec{Y}}$, as done in the above model-based algorithm, using \acp{dnn}: 
	
{\bf End-to-end networks}: Arguably the most common framework is to replace the algorithm with some standard \ac{dnn} architecture. A-priori knowledge of the underlying model is accounted for in the selection of the network type, which is typically treated as a black box. By providing the network with sufficient samples from the joint distribution of the input and label, it should be able to tune its parameters to minimize the loss function. If the network is sufficiently parameterized, i.e., deep, its feasible loss-minimizing structure is within close proximity of the optimal mapping from ${\myVec{Y}}$ to $\hat{\myVec{S}}$, which means that a properly trained network can approach the performance of the model-based algorithm.  The main drawback of using end-to-end networks is that learning a large number of parameters requires a large data set to train. Even when a sufficiently large data set is available, the resulting long training period limits the application of this approach in dynamic environments, where the system has to be retrained often. 

\begin{figure}
	\centering
	\includefig{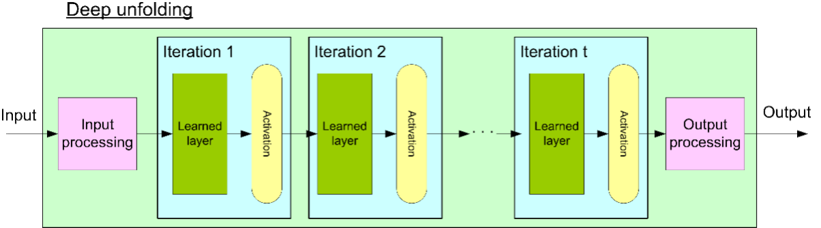} 
	\caption{Deep unfolding illustration.}
	\vspace{-0.4cm}
	\label{fig:Unfolding1}
\end{figure}

{\bf Deep unfolding}: A possible method to design a \ac{dnn} based on an iterative model-based algorithm is to replace each iteration with a layer in the network, where the design of each layer is inspired by the iterative procedure \cite{hershey2014deep,monga2019algorithm}, as illustrated in  Fig. \ref{fig:Unfolding1}. 
	The main benefit of deep unfolding over using end-to-end networks stems from the reduced complexity, as unfolded networks typically have less parameters compared to conventional end-to-end networks \cite{balatsoukas2019deep}. Furthermore, even when the model-based algorithm is feasible, processing $\myVec{Y}$ through a trained unfolded \ac{dnn} is typically faster than applying the iterative algorithm \cite{gregor2010learning}.

\begin{figure}
	\centering
	\includefig{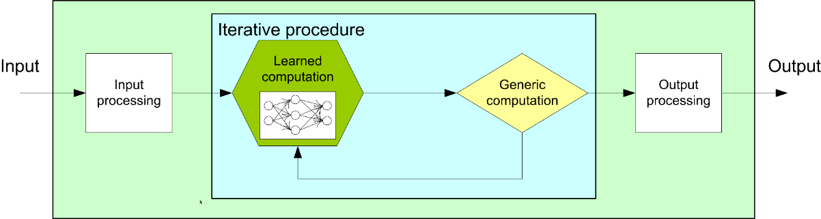} 
	\caption{Integrating \ac{ml} into a model-based algorithm illustration.}
	\label{fig:DNNIntegrate1}
\end{figure}

Here, we propose a new approach  for realizing model-based algorithms of the aforementioned iterative structure in a data-driven fashion. In particular, we first identify the specific computations  which depend on the underlying statistical model $p_{\myVec{S},\myVec{Y}}$. Then we replace these computations with \ac{ml}-based techniques such as dedicated \acp{dnn} that can be trained from data.  In various algorithms, as shown in the following sections, this specific computation often requires accurate knowledge of  $p_{\myVec{S},\myVec{Y}}$ to, e.g., compute a conditional probability measure, or to estimate some parameter, both tasks which can be accurately learned using relatively simple neural networks. An illustration of this strategy is depicted in Fig. \ref{fig:DNNIntegrate1}.

Despite their similarity in combining \acp{dnn} with model-based algorithms, there is a  fundamental difference
between the proposed approach and  unfolding: The main rationale of unfolding is to convert
each iteration of the algorithm into a layer, namely, to design a \ac{dnn} in light of a model-based
algorithm, or alternatively, {\em to integrate the algorithm into the \ac{dnn}}. We propose to implement the  algorithm, while only removing its model-dependence by replacing the model-based computations  with dedicated \acp{dnn}, {\em thus integrating \ac{ml} into the algorithm}.

\vspace{-0,2cm}
\section{Example 1: Finite-Memory Channels}
\label{sec:FiniteMemChan}
\vspace{-0,2cm}
In this section, we present how the rationale for combining \ac{ml} and model-based algorithms detailed in the previous section can be applied for data-driven symbol detection over finite-memory causal channels. We first present the channel model in Section \ref{subsec:ViterbiModel}. Then, we propose the integration of \acp{dnn} into two established detection methods for such channels: We begin with the Viterbi algorithm \cite{forney1973viterbi}, which implements the maximum likelihood sequence detector, proposing a receiver architecture referred to as {\em ViterbiNet}  in Section \ref{subsec:ViterbiDerivation}. ViterbiNet learns to implement the Viterbi algorithm from labeled data. Then, we extend these ideas to propose the {\em BCJRNet} receiver in Section \ref{subsec:BCJRNet}. BCJRNet learns the  factor graph representing the underlying statistical model of the channel, which it uses to implement the symbol-level \ac{map} detector via the BCJR algorithm \cite{bahl1974optimal}. We numerically demonstrate the performance of these \ac{ml}-based receivers in Section \ref{subsec:ViterbiSims}.

\vspace{-0,2cm}
\subsection{System Model}
\label{subsec:ViterbiModel}
\begin{figure}
	\centering
	\includefig{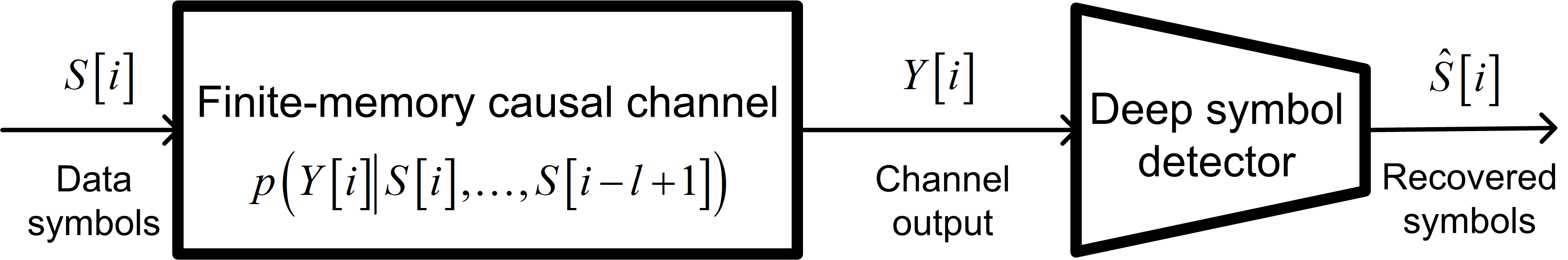} 
	\caption{System model.}
	\label{fig:BasicModel2}
\end{figure}

\vspace{-0,2cm}
We consider the recovery of a block of $\Blklen$ symbols transmitted over a finite-memory stationary causal channel. Let $S[i] \in \mySet{S}$ be the symbol transmitted at time index $i \in \{1,2,\ldots, \Blklen\}\triangleq \Blkset$, where each symbol is uniformly distributed over a set of $\CnstSize$ constellation points, thus $| \mySet{S}| = \CnstSize$. We use $Y[i]\in \mySet{Y}$ to denote the channel output at time index $i$. Since the channel is causal and has a finite memory, $Y[i] $ is given by a stochastic mapping of $\myVec{S}_{i-\Mem+1}^{i}$, where $\Mem$ is the memory of the channel, assumed to be smaller than the blocklength, i.e.,  $\Mem < \Blklen$.
	The conditional \ac{pdf} of the channel output given its input thus satisfies   \eqspace
	\begin{equation}
	\label{eqn:ChModel1}
	\Pdf{\myVec{Y}_{k_1}^{k_2} | \myVec{S}^{ \Blklen}}\left(\myVec{y}_{k_1}^{k_2} | \myVec{s}^{ \Blklen} \right)  = 
	\prod\limits_{i\!=\!k_1}^{k_2}\Pdf{Y[i]  | \myVec{S}_{i\!-\!\Mem\!+\!1}^{i}}\left( y[i]  | \myVec{s}_{i\!-\!\Mem\!+\!1}^{i}\right), 
	\eqspace
	\end{equation}
	for all  $k_1, k_2\in \Blkset$ such that $ k_1 \le k_2$. The fact that the channel is stationary implies that for each $y \in \mySet{Y}$, $\myVec{s} \in \mySet{S}^\Mem$, the conditional \ac{pdf} $\Pdf{Y[i]  | \myVec{S}_{i-\Mem+1}^{i}}\left( y  | \myVec{s}\right)$ does not depend on the  index $i$. An illustration of the system is depicted in Fig. \ref{fig:BasicModel2}.

We focus on two main symbol detector algorithms designed for the finite state channel model in \eqref{eqn:ChModel1}: the Viterbi detector \cite{viterbi1967error}, and the BCJR detector \cite{bahl1974optimal}. The Viterbi scheme implements maximum likelihood sequence detection given by
\eqspace
\begin{align}
\hat{\myVec{s}}^{ \Blklen}\left( \myVec{y}^{ \Blklen}\right)  
&\triangleq \mathop{\arg \max}_{\myVec{s}^{ \Blklen} \in \mySet{S}^\Blklen } \Pdf{\myVec{Y}^{ \Blklen} | \myVec{S}^{ \Blklen}}\left( {\myVec{y}^{ \Blklen} | \myVec{s}^{ \Blklen}}\right)  
= \mathop{\arg \min}_{\myVec{s}^{ \Blklen} \in \mySet{S}^\Blklen } -\log \Pdf{\myVec{Y}^{ \Blklen} | \myVec{S}^{ \Blklen}}\left( \myVec{y}^{ \Blklen} | \myVec{s}^{ \Blklen}\right) .
\label{eqn:ML1}
\eqspace
\end{align}
The BCJR receiver implements the \ac{map} symbol detector  given by
\eqspace
\begin{align}
\hat{S}[k] &= \mathop{\arg \max}\limits_{s \in \mySet{S}}\Pdf{S[k]|\myVec{Y}^t}(s|\myVec{y}^t) 
= \mathop{\arg \max}\limits_{s \in \mySet{S}}\Pdf{S[k],\myVec{Y}^t}(s,\myVec{y}^t), \qquad k \in \Blkset .
\eqspace
\label{eqn:MAP}
\end{align} 
Over the next two sections we present ViterbiNet and BCJRNet, which are two algorithms that learn to carry out Viterbi and BCJR detection, respectively, in a data-driven manner, using \ac{ml} tools.

\vspace{-0,2cm}	
\subsection{ViterbiNet}
\label{subsec:ViterbiDerivation}
\vspace{-0,2cm}
First we present a data-driven framework for recovering  $\myVec{S}^{ \Blklen}$ from the channel output $\myVec{Y}^{ \Blklen}$ based on maximum-likelihood sequence detection.  In particular, in our model the receiver assumes that the channel is stationary, causal, and has finite memory $\Mem$, namely, that the input-output statistical relationship is of the form \eqref{eqn:ChModel1}. The receiver also knows the constellation $\mySet{S}$. We do not assume that the receiver knows the conditional \ac{pdf} $\Pdf{Y[i]  | \myVec{S}_{i\!-\!\Mem\!+\!1}^{i}}\left( y  | \myVec{s}\right)$, i.e., the receiver does not have this \ac{csi}. 
As a preliminary step to designing ViterbiNet, we review Viterbi detection.

\subsubsection{The Viterbi Algorithm} \label{subsec:Viterbi}
The following description of the Viterbi algorithm is based on \cite[Ch. 3.4]{tse2005fundamentals}. 
Since the constellation points are equiprobable, the optimal decision rule in the sense of minimal probability of error is the maximum likelihood decision rule in \eqref{eqn:ML1}.
By defining  
\eqspace
\begin{equation}
c_i\left( \myVec{s}\right)  \triangleq- \log \Pdf{\myVec{Y}[i]  | \myVec{S}_{i-\Mem+1}^{i}}\left( \myVec{y}[i]  | \myVec{s}\right), \qquad \myVec{s} \in \mySet{S}^{\Mem}, 
\label{eqn:CondProb1}
\eqspace
\end{equation} 
it follows from \eqref{eqn:ChModel1} that the log-likelihood in \eqref{eqn:ML1} can be written as 
$\log \Pdf{\myVec{Y}^{ \Blklen} | \myVec{S}^{ \Blklen}} \left( \myVec{y}^{ \Blklen} | \myVec{s}^{ \Blklen}\right) 
= \sum_{i=1}^{\Blklen } c_i\left( \myVec{s}_{i-\Mem+1}^{i}\right)$,
and the optimization problem \eqref{eqn:ML1} becomes
\eqspace
\begin{align}
\hat{\myVec{s}}^{ \Blklen}\left( \myVec{y}^{ \Blklen}\right)   
&= \mathop{\arg \min}_{\myVec{s}^{ \Blklen} \in \mySet{S}^\Blklen }\sum\limits_{i=1}^{\Blklen } c_i\left( \myVec{s}_{i-\Mem+1}^{i}\right).
\label{eqn:ML3}
\eqspace
\end{align}

The optimization problem \eqref{eqn:ML3} can be solved recursively using dynamic programming, by treating the possible combinations of transmitted symbols at each time instance as {\em states} and iteratively updating a {\em path cost} for each state, The resulting scheme, known as the Viterbi algorithm, is given below as Algortihm~\ref{alg:Algo1}. 
\begin{algorithm}
	\caption{ The Viterbi Algorithm \cite{viterbi1967error}}
	\label{alg:Algo1}
	\begin{algorithmic}[1]
		\STATE \underline{Input}: Block of channel outputs $\myVec{y}^{ \Blklen}$, where $\Blklen > \Mem$. 
		\STATE \underline{Initialization}: Set $k\!=\!1$, and fix an initial path cost $\tilde{c}_{0}\!\left( \tilde{\myVec{s}}\right)\! =\!  0$, for each  $\tilde{\myVec{s}} \in \mySet{S}^\Mem$.
		\STATE \label{stp:MF1} For each state $\tilde{\myVec{s}} \in \mySet{S}^\Mem$, compute 
	$	\tilde{c}_k\left( \tilde{\myVec{s}}\right) = \mathop{\min}\limits_{\myVec{u} \in  \mySet{S}^\Mem: \myVec{u}_2^{\Mem} =  \tilde{\myVec{s}}^{\Mem - 1}} \left(\tilde{c}_{k-1}\left( \myVec{u}\right) + {c}_k\left( \tilde{\myVec{s}}\right) \right)$.  
		\STATE \revision{If $k \ge \Mem $, set $\left( \hat{\myVec{s}}\right)_{k\! -\! \Mem\! +\! 1} \!:=\! \left( \tilde{\myVec{s}}_k^{\rm o}\right)_1$, where 
			$\tilde{\myVec{s}}^{\rm o}_k \!=\! \mathop{\arg \min}\limits_{\tilde{\myVec{s}} \in \mySet{S}^\Mem}\tilde{c}_k\left( \tilde{\myVec{s}}\right)$. }  
%
		\STATE Set $k := k+1$. If $k \le \Blklen $ go to Step \ref{stp:MF1}.	
		
		\STATE  \underline{Output}: decoded output $\hat{\myVec{s}}^{ \Blklen}$,  where	\revision{$\hat{\myVec{s}}_{\Blklen-\Mem + 1}^{ \Blklen} := \tilde{\myVec{s}}_{{ \Blklen}}^{\rm o}$.}
	\end{algorithmic}
\end{algorithm} 

The Viterbi algorithm has two major advantages: $1)$ The algorithm solves \eqref{eqn:ML1} at a computational complexity that is linear in the blocklength $\Blklen$. For comparison, the computational complexity of solving \eqref{eqn:ML1} directly grows exponentially with $\Blklen$;   	
	$2)$ The algorithm produces estimates sequentially during run-time. In particular, while in \eqref{eqn:ML1}  the estimated output $\hat{\myVec{s}}^{ \Blklen}$ is computed using the entire received block $\myVec{y}^{ \Blklen}$, Algorithm~\ref{alg:Algo1} computes $\hat{s}[i]$  once $y[i + \Mem - 1]$ is received.  

In order to implement Algorithm \ref{alg:Algo1}, one must  compute  ${c}_i\left( {\myVec{s}}\right)$ of \eqref{eqn:CondProb1} for all $i \in \Blkset$ and for each $\myVec{s} \in \mySet{S}^\Mem$. Consequently, the conditional \ac{pdf} of the channel, which we refer to as full \ac{csi}, must be explicitly known. As discussed in the introduction, obtaining full \ac{csi} may be extremely difficult in rapidly changing channels and may also require a large training overhead. 
In the following we propose ViterbiNet,  an \ac{ml}-based detector based on the Viterbi algorithm, that does not require \ac{csi}. 

\subsubsection{Integrating \ac{ml} into the Viterbi Algorithm}
\label{subsubsec:ViterbiNetDerivation}
\vspace{-0.1cm}
%
%
%
In order to integrate \ac{ml} into the Viterbi Algorithm, we note that  \ac{csi} is  required in Algorithm~\ref{alg:Algo1} only in Step~\ref{stp:MF1} to compute the log-likelihood function $ c_i (\myVec{s})$. 
Once  $ c_i (\myVec{s})$ is computed for each $\myVec{s} \in \mySet{S}^\Mem$, the Viterbi algorithm only requires knowledge of the memory length $\Mem$ of the channel. This requirement is much easier to satisfy compared to full \ac{csi} (e.g., by using an upper bound on memory length). 

\begin{figure}
	\centering
	\includegraphics[width = 0.5\columnwidth]{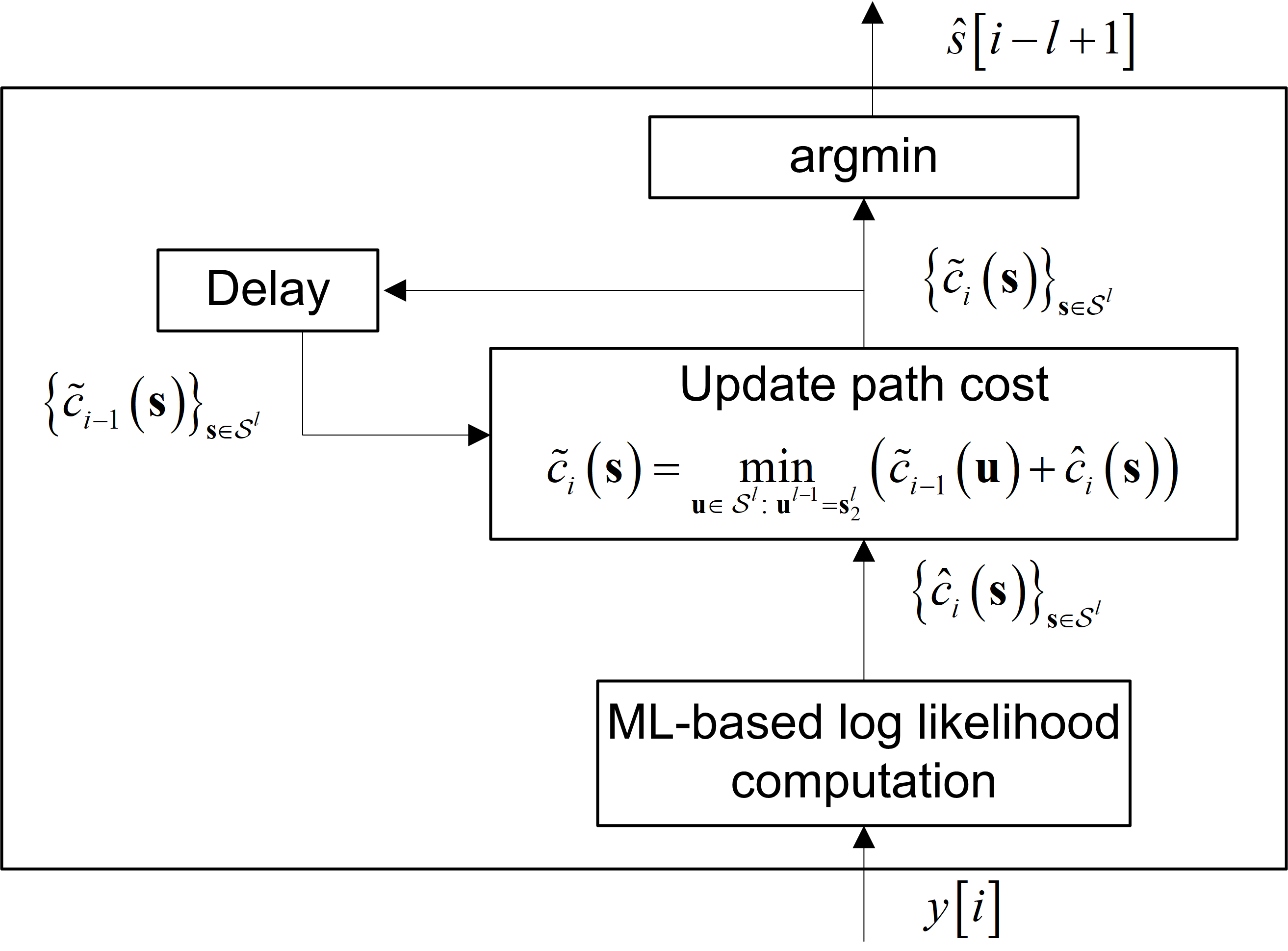} 
	\vspace{-0.2cm}
	\caption{ViterbiNet Architecture.}
	\label{fig:DNNSystem}
\end{figure}

Since the channel is stationary, it holds by \eqref{eqn:CondProb1} that if $y[i] = y[k]$ then $ c_i (\myVec{s}) = c_k(\myVec{s})$, for each $\myVec{s} \in \mySet{S}^\Mem$, and  $i,k \in \Blkset$. Consequently, the log-likelihood function  $ c_i (\myVec{s})$ depends only on the values of $y[i]$ and of $\myVec{s}$, and not on the time index $i$. 
Therefore, to implement Algorithm \ref{alg:Algo1} in a data-driven fashion, we replace the explicit computation of the log-likelihood \eqref{eqn:CondProb1} with an \ac{ml}-based system that learns to evaluate the cost function from the training data. In this case, the input of the system is $y[i]$ and the output is an estimate of $c_i({\myVec{s}})$, denoted $\hat{c}_i({\myVec{s}})$, for each $\myVec{s}\in\mySet{S}^{\Mem}$. The rest of the Viterbi algorithm remains intact, and the detector implements Algorithm~\ref{alg:Algo1} using the learned $c_i({\myVec{s}})$.   The proposed architecture is illustrated in Fig.~\ref{fig:DNNSystem}.

A major challenge in implementing a network capable of computing $c_i(\myVec{s})$ from $y[i]$ stems from the fact that, by \eqref{eqn:CondProb1}, $c_i(\myVec{s})$ represents the log-likelihood of $Y[i] = y[i]$ given $\myVec{S}_{i-\Mem+1}^i = \myVec{s}$. However,  \acp{dnn} trained with the cross entropy loss typically output the conditional distribution of $\myVec{S}_{i-\Mem+1}^i = \myVec{s}$ given  $Y[i] = y[i]$, i.e., $\Pdf{\myVec{S}_{i-\Mem+1}^i| Y[i]}\left( \myVec{s}| y[i]\right)$. Specifically, classification \acp{dnn} with input $Y[i] = y[i]$, which typically takes values in a discrete set, or alternatively, is discretized by binning a continuous quantity as in  \cite{oord2016wavenet}, 
output the  distribution of the label $\myVec{S}_{i-\Mem+1}^i$ conditioned on that input $Y[i] = y[i]$, and not the distribution of the input conditioned on all possible values of the label  $\myVec{S}_{i-\Mem+1}^i$.
For this reason, the previous work \cite{farsad2018neural} used a \ac{dnn} to approximate the \ac{map} detector by considering  $\Pdf{S[i]| \myVec{Y}_{i-\Mem+1}^i}\left( s| \myVec{y}_{i-\Mem+1}^i\right)$.
The quantity needed for the Viterbi algorithm is the conditional \ac{pdf} $\Pdf{Y[i]|\myVec{S}_{i\!-\!\Mem\!+\!1}^i}\!\left(  y[i]| \myVec{s}\right)$, and not the conditional distribution $\Pdf{\myVec{S}_{i\!-\!\Mem\!+\!1}^i| Y[i]}\!\left( \myVec{s}| y[i]\right)$.  
Note that while $\sum_{\myVec{s}\in\mySet{S}^\Mem} \Pdf{\myVec{S}_{i-\Mem+1}^i| Y[i]}\left( \myVec{s}| y[i]\right) = 1$, for the desired conditional \ac{pdf}  $\sum_{\myVec{s}\in\mySet{S}^\Mem} \Pdf{Y[i]|\myVec{S}_{i-\Mem+1}^i}\left(  y[i]| \myVec{s}\right) \neq 1$ in general.  
Therefore, outputs generated using conventional \acp{dnn} with a softmax output layer are not applicable. 
The fact that  Algorithm \ref{alg:Algo1}  uses $\Pdf{Y[i]|\myVec{S}_{i-\Mem+1}^i}\left(  y[i]| \myVec{s}\right)$ instead of $\Pdf{\myVec{S}_{i-\Mem+1}^i| Y[i]}\left( \myVec{s}| y[i]\right)$ allows it to exploit the Markovian nature of the channel, induced by the finite memory in \eqref{eqn:ChModel1}, resulting in the aforementioned advantages of the Viterbi algorithm.

%

To tackle this difficulty, we recall that by Bayes' theorem, as the symbols are equiprobable, the desired conditional \ac{pdf} $\Pdf{Y[i]|\myVec{S}_{i-\Mem+1}^i}\left(  y| \myVec{s}\right)$ can be written as 
\eqspace
\begin{equation}
\label{eqn:Bayes}
\Pdf{Y[i]|\myVec{S}_{i-\Mem+1}^i}\left(  y| \myVec{s}\right) = { \CnstSize^{\Mem}}\cdot{\Pdf{\myVec{S}_{i-\Mem+1}^i| Y[i]}\left( \myVec{s}| y\right)\cdot\Pdf{Y[i]}(y)}.
\eqspace
\end{equation} 
Therefore, given estimates of  $\Pdf{Y[i]}(y[i])$ and of  $\Pdf{\myVec{S}_{i-\Mem+1}^i| Y[i]}\left( \myVec{s}| y[i]\right)$ for each $\myVec{s} \in \mySet{S}^\Mem$, the log-likelihood function $c_i(\myVec{s})$ can be recovered using \eqref{eqn:Bayes} and \eqref{eqn:CondProb1}.

\begin{figure}
		\centering
		\includefig{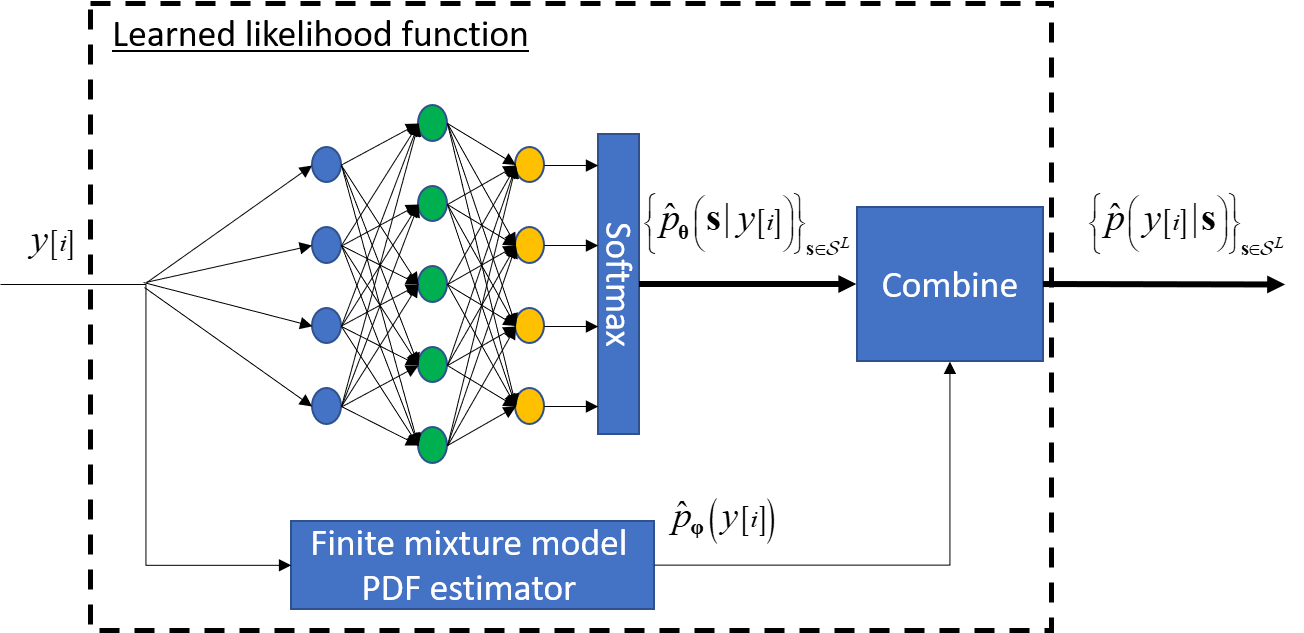}
		\caption{\ac{ml}-based likelihood computation.}
		\label{fig:NetworkArchitecture}
	\end{figure}

A parametric estimate of $\Pdf{\myVec{S}_{i-\Mem+1}^i| Y[i]}\left( \myVec{s}| y[i]\right)$, denoted  $\PdfEst{\myVec{\theta}}\left( \myVec{s}| y[i]\right)$, can be reliably obtained from training data using standard classification \acp{dnn}  with a softmax output layer. The marginal \ac{pdf} of $Y[i]$ may be estimated from the training data using conventional kernel density estimation methods. Furthermore, the fact that $Y[i]$ is a stochastic mapping of $\myVec{S}_{i-\Mem+1}^i$ implies that its distribution can be approximated as a mixture model of $\CnstSize^\Mem$ kernel functions \cite{mclachlan2004finite}. Consequently, a parametric estimate of $\Pdf{Y[i]}(y[i])$, denoted $\PdfEst{\myVec{\varphi}}\left(  y[i]\right)$, can be obtained from the training data using mixture density networks \cite{bishop1994mixture}, \ac{em}-based algorithms \cite[Ch. 2]{mclachlan2004finite}, or any other finite mixture model fitting method. The resulting \ac{ml}-based log-likelihood computation is illustrated in Fig. \ref{fig:NetworkArchitecture}.

\vspace{-0,2cm}
\subsection{BCJRNet}
\label{subsec:BCJRNet}
\vspace{-0,2cm}
In this section, we expand the approach presented in the previous section to factor graphs and  detail how this strategy can be used to implement BCJRNet, which is a data-driven channel-model-independent \ac{map} symbol detector for finite-memory channels. We first discuss how \ac{ml} can be incorporated into factor graphs, and then present the BCJRNet detector.

\subsubsection{Learning Factor Graphs}
\label{subsec:FGLearn} 
Factor graphs are graphical models that are used to represent factorization of multivariate functions (e.g. joint probability distributions) \cite{kschischang2001factor}. In factor graphs, each factor is represented by a {\em factor or function node}, and each variable is represented by an edge or half-edge in the graph\footnote{In this work we use the style proposed in \cite{forney2001codes}, known as {\em Forney-style factor graphs.}}. Factor graph methods, such as the sum-product algorithm \cite{kschischang2001factor}, exploit the factorization of a joint distribution to efficiently compute a desired quantity by passing messages along the edges of the graph. For example, the application of the sum-product algorithm in the finite-memory channel in \eqref{eqn:ChModel1} exploits its factorization to compute marginal distributions, an operation whose burden typically grows exponentially with the block size, with complexity that only grows linearly with $t$. In fact, the sum-product algorithm is exactly the recursive computation carried out in the BCJR symbol detector \cite{bahl1974optimal}. More generally, the sum-product algorithm specializes a multitude of common signal processing techniques, including the Kalman filter and \ac{hmm} prediction \cite{loeliger2004introduction}. 

\begin{figure}
	\centering
	{\includefig{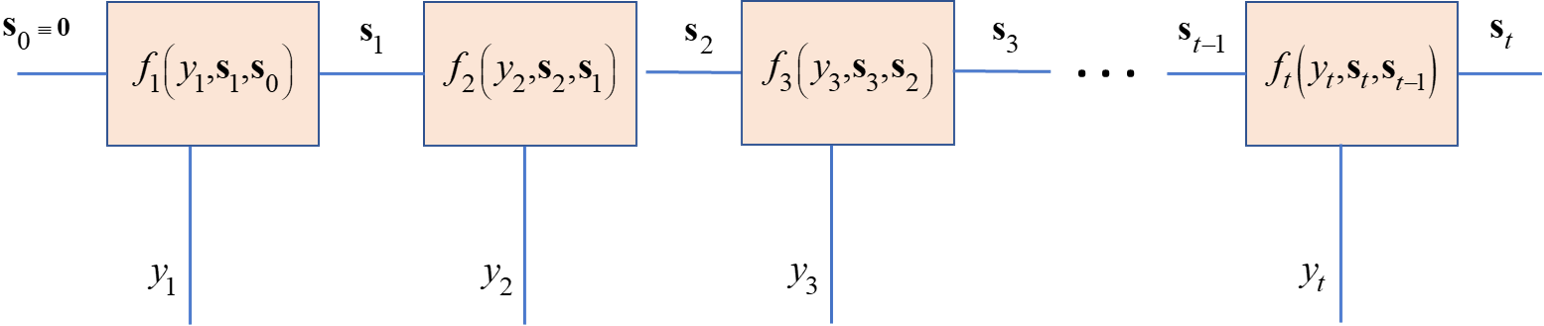}} 
	\caption{Factor graph of finite-memory channel.}
	\label{fig:FGFinite}	 
\end{figure}

The factor graph representing finite-memory channels  \eqref{eqn:ChModel1} is presented in Fig.~\ref{fig:FGFinite}, where we use the simplifying notation $\myVec{s}_i=\myVec{s}_{i-l+1}^{i}$, $y_i=y[i]$, and 
the fact that
\eqspace
\begin{align*}
P_{\myVec{Y}^t, \myVec{S}^t}(\myVec{y}^t, \myVec{s}^t)  
	&\!=\! \prod_{i=1}^{t}	P_{Y[i]| \myVec{S}_{i\!-\!l\!+\!1}^{i}}\!\left( y_i| \myVec{s}_i\right)	P_{\myVec{S}_{i\!-\!l\!+\!1}^{i}| \myVec{S}_{i\!-\!l}^{i\!-\!1}}\!\left( \myVec{s}_{i}| \myVec{s}_{i\!-\!1}\right)  
	\!=\! \prod_{i=1}^{t}	f_i(y_i,\myVec{s}_{i}, \myVec{s}_{i\!-\!1}).
\eqspace
\end{align*}
In order to implement the sum-product scheme, one must be able to specify the factor graph encapsulating the underlying distribution, and in particular, the function nodes $\{f_j\}_{j=1}^{t}$. This implies that the BCJR detector requires full \ac{csi}.

We next generalize the approach we took in the previous section with the ViterbiNet to realize a broad family of data-driven factor graph methods by learning the mappings carried out at the function nodes  from a relatively small set of labeled data using \ac{ml} tools. By doing so, one can train a system to learn an underlying factor graph, which can then be utilized for inference using conventional factor graph methods, such as the sum-product algorithm. 

The proposed approach requires prior knowledge of the graph structure, but not its nodes. For example, a finite-memory channel with memory length not larger than $l$ can be represented using the structure in Fig. \ref{fig:FGFinite} while its specific input-output relationship dictate the function nodes. Consequently, in order to learn such a factor graph from samples, one must learn its function nodes. As these mappings often represent conditional distribution measures, they can be naturally learned using classification networks, e.g., fully-connected \acp{dnn} with a softmax output layer and cross-entropy objective, which are known to reliably learn conditional distributions in complex environments \cite{bengio2009learning}. 
Furthermore, in many scenarios of interest, e.g., stationary finite-memory channels and time-invariant state-space models, the mapping implemented by the factor nodes $f_i(\cdot)$ does not depend on the index $i$. In such cases, only a fixed set of mappings whose size does not grow with the dimensionality $t$ has to be tuned in order to learn the complete factor graph. An example of how this concept of learned factor graphs can be applied to the BCJR detector is presented next.

\subsubsection{BCJRNet: Data-Driven MAP Recovery}
\label{subsec:BCJRDerivation}
\vspace{-0.1cm}
In this section, we first present the application of the sum-product method for \ac{map} symbol detection in finite-memory channels, also known as the BCJR algorithm \cite{bahl1974optimal}. We then describe how the approach proposed in Section \ref{sec:ModelML} can be incorporated into the BCJR algorithm. 

Consider a stationary finite-memory channel, namely, a channel obeying the model in Section \ref{subsec:ViterbiModel}. 
The detector which minimizes the symbol error probability is the \ac{map} rule given in \eqref{eqn:MAP}.
Using the formulation of the state vectors $\{\myVec{S}_{k-l+1}^k\}$, the desired joint probability can be written as \cite[Ch. 9.3]{cioffi2008equalization}
\eqspace
\begin{equation}
\Pdf{S[k],\myVec{Y}^t}(s_k,\myVec{y}^t) \!= \!\!\sum_{\myVec{s}\in\mySet{S}^L} \!\!\Pdf{\myVec{S}_{k-l}^{k-1},\myVec{S}_{k-l+1}^{k}, \myVec{Y}^t}(\myVec{s},[s_k, (\myVec{s})_1 \ldots  (\myVec{s})_{l\!-\!1}]^T\!, \myVec{y}^t).
\label{eqn:Joint1}
\eqspace
\end{equation}
The summands in \eqref{eqn:Joint1} are the joint distributions evaluated recursively from the channel factor graph. Thus, when the factor graph is known, the \ac{map} rule \eqref{eqn:MAP} can be computed efficiently using the sum-product algorithm. 

Now, we show how the rationale presented in Section \ref{sec:ModelML} yields a method for learning the factor graphs of finite-memory channels, from which the \ac{map} detector can be obtained. 
We assume that the channel memory length, $l$, is known. However, the channel model, i.e., the conditional distribution $ \Pdf{Y[i] | \myVec{S}_{i-\Mem+1}^{i}}(\cdot)$ is unknown, and only a set of labeled input-output pairs
is available.  

Since the channel memory is known, the structure of the factor graph is fixed to that depicted in Fig.~\ref{fig:FGFinite}. Consequently, in order to learn the factor graph, one must only adapt the function nodes $f_i(\cdot)$, 
which are given by 
\eqspace
\begin{align}
f_i(y_i,\myVec{s}_{i}, \myVec{s}_{i-1}) 
&= P_{Y[i]| \myVec{S}_{i-l+1}^{i}}\left( y_i| \myVec{s}_i\right)	P_{\myVec{S}_{i-l+1}^{i}| \myVec{S}_{i-l}^{i-1}}\left( \myVec{s}_{i}| \myVec{s}_{i-1}\right) \notag \\
&\stackrel{(a)}{=} 
\begin{cases}
 \frac{1}{m} \Pdf{Y| \myVec{S}}\left( y_i| \myVec{s}_i\right) & (\myVec{s}_i)_j = (\myVec{s}_{i\!-1})_{j\!-\!1}, \quad \forall j\in\{2,\ldots, l\} \\
 0 & {\rm otherwise},
\end{cases}
\label{eqn:FunctionNode}
\eqspace
\end{align}
where $(a)$ follows from the definition of the state vectors $\{\myVec{S}_{k-l+1}^k\}$ and the channel stationarity, which implies that $\Pdf{Y[i] | \myVec{S}_{i-\Mem+1}^{i}}(\cdot)$ does not depend on $i$ and can be written as $\Pdf{Y | \myVec{S}}(\cdot)$. The formulation of the function nodes in \eqref{eqn:FunctionNode} implies that they can be estimated by training an \ac{ml}-based system to evaluate  $\Pdf{Y | \myVec{S}}(\cdot)$ from which the corresponding function node value is obtained via \eqref{eqn:FunctionNode}. Once the factor graph representing the channel is learned, symbol recovery is carried out using the sum-product method detailed earlier. The resulting receiver, referred to as {\em BCJRNet}, thus implements BCJR detection in a data-driven manner, and is expected to approach \ac{map}-performance when the function nodes are accurately estimated, as demonstrated in our numerical study in the next section.

Taking the same approach as ViterbiNet, since $Y[i]=y_i$ is given and may take continuous values while $\myVec{S}_{i-l+1}^{i}$, representing the label, takes discrete values, a natural approach to evaluate $\Pdf{Y[i]|\myVec{S}_{i-l+1}^{i}}(y_i|\myVec{s}_i)$ for  each $\myVec{s}_i\in\mySet{S}^l$ using \ac{ml} tools is to estimate $\Pdf{\myVec{S}_{i-l+1}^{i}|Y[i]}(\myVec{s}_i|y_i)$, from which the desired $\Pdf{Y[i]|\myVec{S}_{i-l+1}^{i}}(y_i|\myVec{s}_i)$  can be obtained using Bayes rule via \eqref{eqn:Bayes}.
In particular, BCJRNet utilizes two parametric models: one for evaluating the conditional  $\Pdf{\myVec{S}_{i-l+1}^{i}|Y[i]}(\myVec{s}_i|y_i)$, and another for computing the marginal \ac{pdf} $\Pdf{Y[i]}(y_i)$, using a similar system as that used for computing the likelihood functions in ViterbiNet, i.e., via the architecture illustrated in Fig. \ref{fig:NetworkArchitecture}. The estimates are then combined into the learned function nodes using \eqref{eqn:FunctionNode} and \eqref{eqn:Bayes}, which are used by BCJRNet to carry out BCJR detection in a data-driven manner.


\vspace{-0,2cm}
\subsection{Numerical Evaluations}
\label{subsec:ViterbiSims}
\vspace{-0,2cm}
In this section, we numerically compare the performance of the proposed data-driven ViterbiNet and BCJRNet to the conventional model-based Viterbi algorithm and BCJR detector as well as to previously proposed deep symbol detectors. 
	Throughout this numerical study we implement the fully-connected network in Fig. \ref{fig:NetworkArchitecture} using three layers: a $1 \times 100$ layer followed by a $100 \times 50$ layer and a $50 \times  16 (=|\CnstSize|^{\Mem})$ layer, using intermediate sigmoid and ReLU activation functions, respectively. The mixture model estimator approximates the distribution as a Gaussian mixture using \ac{em}-based fitting \cite[Ch. 2]{mclachlan2004finite}. \revision{The network is trained using $5000$ training samples to minimize the cross-entropy loss via the Adam optimizer \cite{kingma2014adam} with learning rate $0.01$, using up to $100$ epochs with mini-batch size of $27$ observations.} We note that the number of training samples is of the same order and even smaller compared to typical preamble sequences in wireless communications. Due to the small number of training samples and the simple architecture of the \ac{dnn}, only a few minutes are required to train the network on a standard CPU.
	
		 We consider two finite-memory  channels: An \ac{isi} channel with \ac{awgn}, and a Poisson channel. In both channels we set the memory length to $\Mem = 4$. 
	 For the  \ac{awgn} channel, we let $W[i]$ be a zero-mean unit variance \ac{awgn} independent of $S[i]$, and let $\myVec{h} (\gamma)\in \mySet{R}^\Mem$ be the channel vector obeying an exponentially decaying profile  $\left( \myVec{h}\right)_\tau \triangleq e^{-\gamma(\tau-1)}$ for $\gamma > 0$. The  input-output relationship is given by
\eqspace
	 \begin{equation}
	 \label{eqn:AWGNCh1}
	 Y[i] = \sqrt{\rho} \cdot\sum\limits_{\tau=1}^{\Mem} \left( \myVec{h}(\gamma)\right)_\tau S[i-\tau + 1] + W[i],
\eqspace
	 \end{equation}
	 where $\rho > 0$ represents the \ac{snr}.  
	 The channel input is randomized from a \ac{bpsk} constellation, i.e., $\mySet{S} = \{-1, 1\}$. 
	 For the Poisson channel, the channel input represents on-off keying, namely, $\mySet{S} = \{0,1\}$, and the channel output $Y[i]$ is generated from the input via 
\eqspace
	 \begin{equation}
	\label{eqn:PoissonCh1}
	Y[i] | \myVec{S}^\Blklen\sim \mathds{P}\left( \sqrt{\rho} \cdot\sum\limits_{\tau=1}^{\Mem} \left( \myVec{h}(\gamma)\right)_\tau S[i-\tau + 1] + 1\right),
\eqspace
	 \end{equation}
	 where $\mathds{P}(\lambda)$ is the Poisson distribution with parameter $\lambda > 0$.
	
 	For each channel, we numerically compute the \ac{ser} of  ViterbiNet and BCJRNet for  different values of the \ac{snr} parameter $\rho$. In the following study, the \ac{dnn} in Fig, \ref{fig:NetworkArchitecture}, which produces a parametric estimate of the log-likelihoods in ViterbiNet and produces the learned function nodes in BCJRNet, is trained anew for each value of $\rho$.
	For each \ac{snr} $\rho$, the \ac{ser} values are averaged over $20$ different channel vectors  $\myVec{h} (\gamma)$, obtained by letting $\gamma$ vary in the range $[0.1, 2]$. 
	 For comparison, we  numerically compute the \ac{ser} of the Viterbi algorithm, as well as that of the \ac{sbrnn} deep symbol decoder proposed in \cite{farsad2018neural}. 
	 \label{txt:Robustness1}
	 In order to study the resiliency of the data-driven detectors to inaccurate training, we also compute the performance when the receiver only has access to a noisy estimate of $\myVec{h}(\gamma)$, and specifically, to a copy of $\myVec{h}(\gamma)$ whose entries are corrupted by i.i.d. zero-mean Gaussian noise with variance $\sigma_e^2$. In particular, we use  $\sigma_e^2 = 0.1$ for the Gaussian channel \eqref{eqn:AWGNCh1}, and $\sigma_e^2 = 0.08$ for the Poisson channel \eqref{eqn:PoissonCh1}. 
	We consider two cases: {\em Perfect \ac{csi}}, in which the channel-model-based detectors have accurate knowledge of  $\myVec{h}(\gamma)$, while the data-driven receivers are trained using labeled samples generated with the same  $\myVec{h}(\gamma)$ used for generating the test data; and {\em \ac{csi} uncertainty}, where the model-based algorithms are implemented with the log-likelihoods (for Viterbi algorithm) and function nodes (for BCJR detection) computed using the noisy version of  $\myVec{h}(\gamma)$,  while the  data used for training ViterbiNet and BCJRNet is generated with  the noisy version of $\myVec{h}(\gamma)$ instead of the true one. 
	In all cases, the information symbols are uniformly randomized in an i.i.d. fashion from $\mySet{S}$, and the test samples are generated from their corresponding channel
	with the true  channel vector $\myVec{h}(\gamma)$. 
   
\begin{figure}
	\centering
	\begin{minipage}{0.45\textwidth}
		\centering
		\scalebox{0.38}{\includegraphics{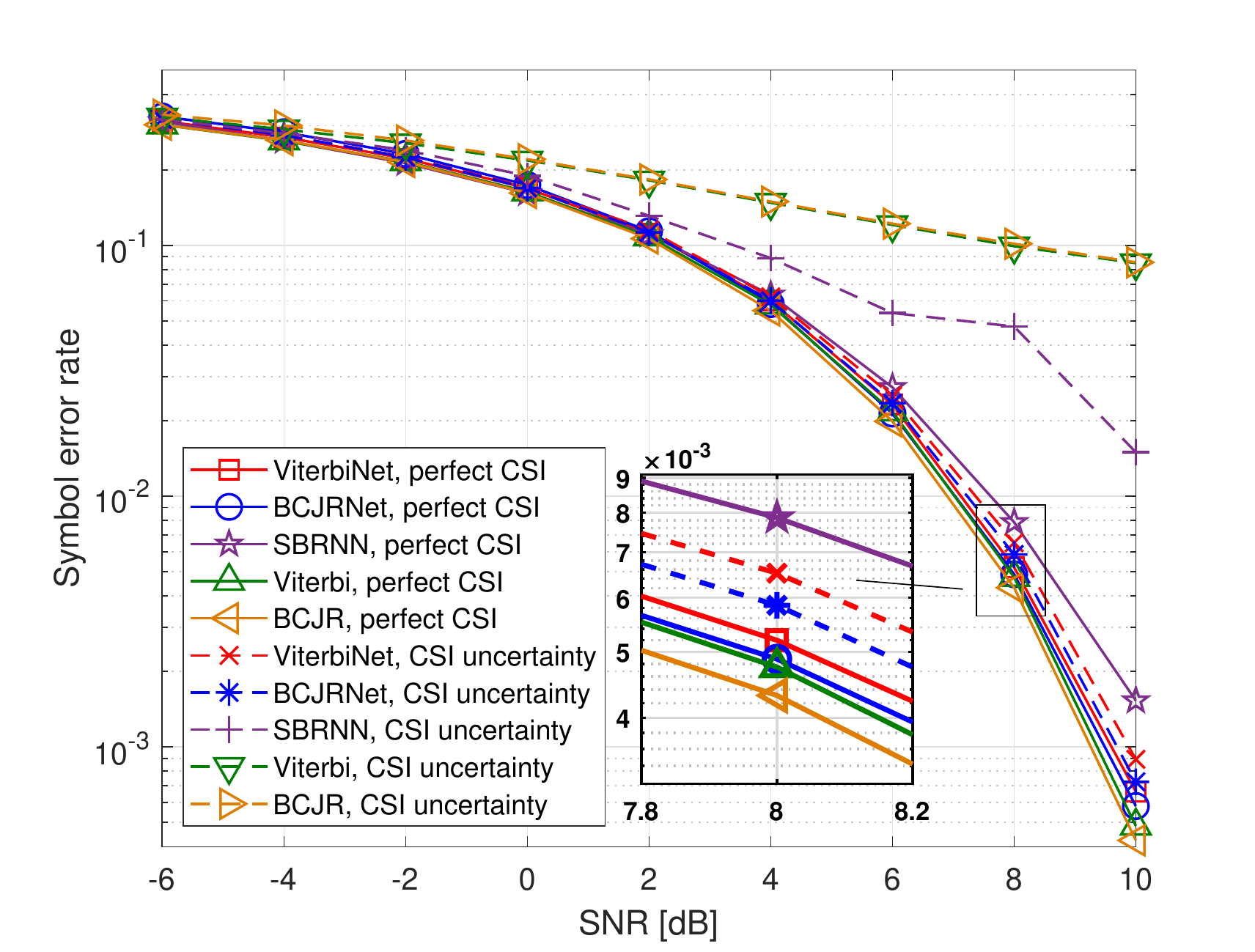}}
		\figSpace
		\caption{\ac{isi} channel with \ac{awgn}.
		}
		\label{fig:AWGN} 	
	\end{minipage}
	$\quad$
	\begin{minipage}{0.45\textwidth}
		\centering
		\scalebox{0.38}{\includegraphics{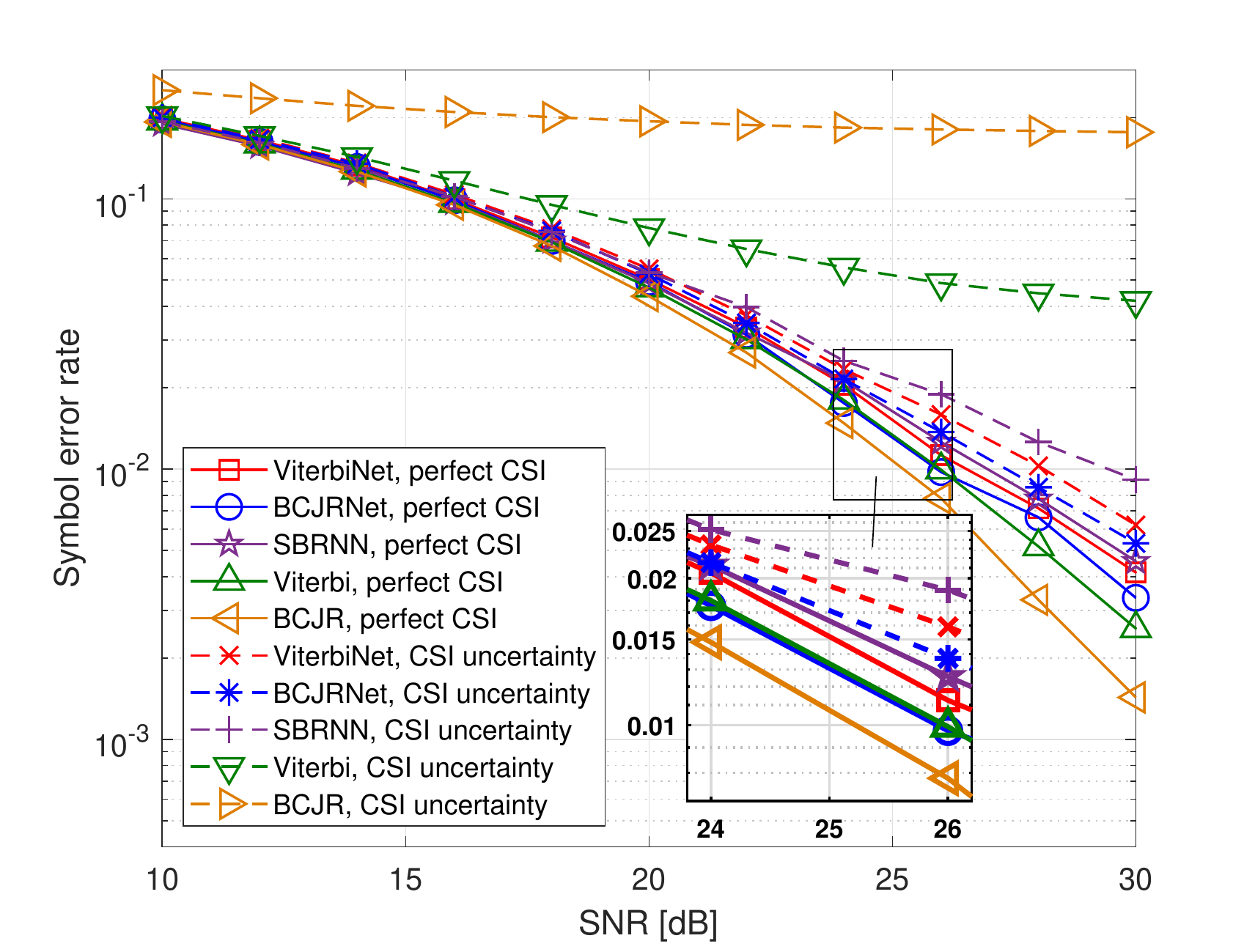}}
    	\figSpace
    	\caption{Poisson channel.
    	}
    	\label{fig:Poisson} 
	\end{minipage}
\end{figure}

	The numerically computed \ac{ser} values, averaged over $50000$ Monte Carlo simulations, versus $\rho \in [-6,10]$ dB for the \ac{isi} channel with \ac{awgn} are depicted in Fig. \ref{fig:AWGN}, while the corresponding performance versus $\rho \in [10,30]$ dB for the Poisson channel are depicted in Fig. \ref{fig:Poisson}. Observing Figs. \ref{fig:AWGN}-\ref{fig:Poisson}, we note that the performance of the data-driven receivers approaches that of their corresponding \ac{csi}-based counterparts. 
	In particular, for the \ac{awgn} case, in which the channel output obeys a Gaussian mixture distribution, the performance of ViterbiNet coincides with that of the Viterbi algorithm, while the \ac{ser} of BCJRNet is within a very small gap of the \ac{csi}-based BCJR receiver. For the Poisson channel, both data-driven receivers achieve performance within a small gap of the model-based ones, which is more notable at high \acp{snr}. This gap stems from the model mismatch induced by approximating the distribution of $Y[i]$ as a Gaussian mixture. 
We also observe that the \ac{sbrnn} receiver, which was shown in \cite{farsad2018neural} to approach the performance of the \ac{csi}-based Viterbi algorithm when sufficient training is provided, is outperformed by ViterbiNet and BCJRNet here due to the small training set size. 
	These results demonstrate that our proposed data-driven detectors, which uses simple \ac{dnn} structures embedded into an established detection algorithms, require significantly less training compared to previously proposed \ac{ml}-based receivers.
	
	In the presence of \ac{csi} uncertainty, it is observed in Figs. \ref{fig:AWGN}-\ref{fig:Poisson} that both ViterbiNet and BCJRNet significantly outperform the model-based algorithms from which they originate. In particular, when ViterbiNet and BCJRNet are trained with a variety of different channel conditions, they are still capable of achieving relatively good \ac{ser} performance under each of the channel conditions for which it is trained, while the performance of the conventional Viterbi and BCJR algorithms is significantly degraded in the presence of imperfect \ac{csi}. While the \ac{sbrnn} receiver is shown to be more resilient to inaccurate \ac{csi} compared to the Viterbi and BCJR algorithms, as was also observed in \cite{farsad2018neural}, it is outperformed by  ViterbiNet  and BCJRNet with the same level of uncertainty, and the performance gap is more notable in the \ac{awgn} channel. The reduced gain of ViterbiNet and BCJRNet over the \ac{sbrnn} receiver for the Poisson channel stems from the fact that the \ac{ml}-based module of Fig. \ref{fig:NetworkArchitecture}  uses a Gaussian mixture  density estimator for the \ac{pdf} of $Y[i]$, which obeys a Poisson mixture distribution for the  channel \eqref{eqn:PoissonCh1}. 
	
The results presented in this section demonstrate that \ac{ml}-based receivers can be trained to carry out accurate and robust  symbol detection using a relatively small amount of labeled samples. This is achieved by utilizing dedicated \ac{ml} tools to learn only the model-based computations of symbol detection algorithms, such as the Viterbi and BCJR methods, while cobmining  these learned computations with the generic operations of the algorithms, e.g., the dynamic programming processing of the Viterbi algorithm or the sum-product recursion of the BCJR receiver.

\vspace{-0,2cm}
\section{Example 2: Memoryless MIMO Channels}
\label{sec:DeepSIC}
\vspace{-0,2cm}
In the previous section we utilized the rationale detailed in Section \ref{sec:ModelML} for combining \ac{ml} and model-based algorithms in order to implement two symbol detection schemes - the Vitebi detector and the BCJR algorithm - in a data-driven fashion. Both these methods are based on computing the likelihood for each possible combination of channel inputs which could have lead to the observed output, i.e., for each state, followed by some generic recursive operations. Such methods, both the original channel-model-based algorithms as well as their data-driven \ac{ml}-based implementations, grow computationally infeasible when multiple symbols are transmitted simultaneously, as in, e.g., multiuser \ac{mimo} systems, since the possible number of states grows rapidly with the number of transmitted symbols. 
In this section we demonstrate how the proposed model-based \ac{ml} rationale can be exploited to realize symbol detection methods suitable for such \ac{mimo} systems in a data-driven manner. We begin by presenting the channel model in Section \ref{subsec:DeepSICModel}. Then, in Section \ref{subsec:DeepSICDerivation} we show how the iterative \ac{sic} symbol detection mapping proposed in \cite{choi2000iterative}, which is a  \ac{mimo} receiver scheme capable of approaching \ac{map} performance at affordable complexity, can be learned from training, resulting in an \ac{ml}-based receiver referred to as {\em DeepSIC}. The achievable performance of DeepSIC is demonstrated in a numerical study presented in Section \ref{subsec:DeepSICSims}. 

\vspace{-0,2cm}
\subsection{System Model}
\label{subsec:DeepSICModel}
\begin{figure}
	\centering
	\includefig{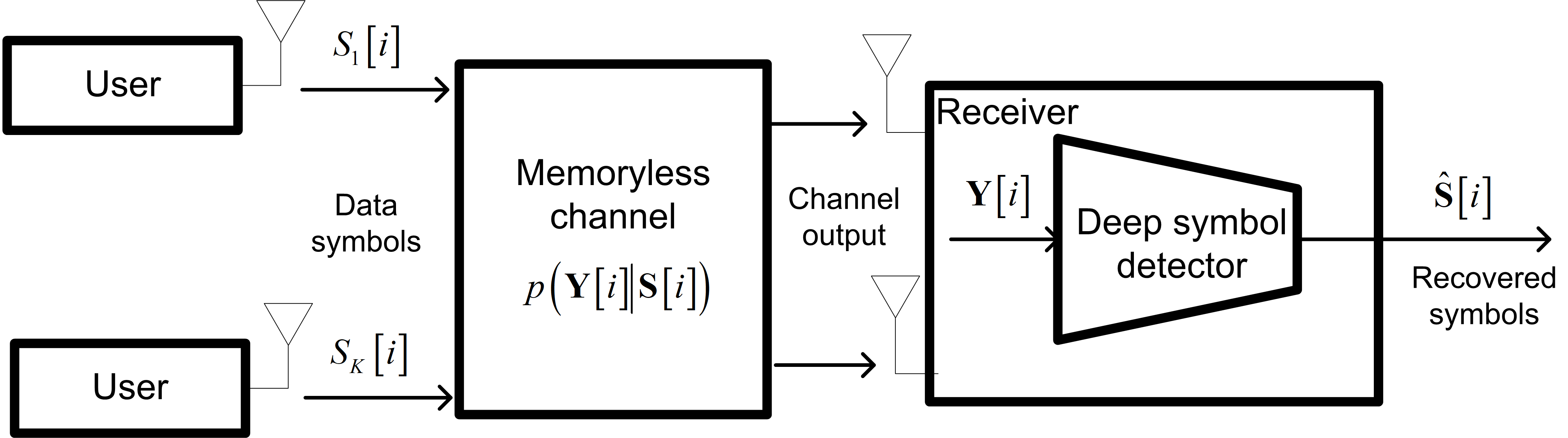} 
	\caption{System model.}
	\label{fig:ChannelModel2}
\end{figure}

\vspace{-0,2cm}
	We consider a multiuser uplink \ac{mimo} system in which $\Nusers$ single antenna users communicate with a receiver equipped with $\Nantennas$ antennas over a memoryless stationary channel.  
	At each time instance $i$, the $k$th user, $k \in \{1,2,\ldots, \Nusers\}\triangleq \NusersSet$, transmits a symbol $S_k[i]$  drawn from a constellation $\mySet{S}$ of size $\CnstSize$. Each symbol is uniformly distributed over $\mySet{S}$, and the symbols transmitted by different users are mutually independent.   We use $\myY[i]\in \mySet{R}^{\Nantennas}$ to denote the channel output at time index $i$. While we focus on real-valued channels, the system model can be adapted to complex-valued channels, as complex vectors can be equivalently represented using real vectors of extended dimensions.  Since the channel is memoryless, $\myY[i] $ is given by some stochastic mapping of $\myS[i] \triangleq \big[S_1[i], S_2[i], \ldots, S_{\Nusers}[i]\big]^T$, represented by the conditional distribution  $\Pdf{\myY[i] | \myS[i]}$.  Since the channel is stationary, this conditional distribution does not depend on the  index $i$, and is thus denoted henceforth by $\Pdf{\myY | \myS}$. An illustration of the system is depicted in Fig. \ref{fig:ChannelModel2}.

	We focus on the problem of recovering the transmitted symbols  $\myS[i]$ from the channel output $\myY[i]$. 
	The optimal detection rule which minimizes the probability of error given a channel output realization $\myY[i] = \myVec{y}$ is the \ac{map} detector. Letting $\Pdf{\myS | \myY}$ be the conditional distribution of $\myS[i]$ given $\myY[i]$, the \ac{map} rule is given by
\eqspace
	\begin{equation}
	\label{eqn:MAPMimo}
	\hat{\myVec{s}}_{\rm MAP}[i] \triangleq \mathop{\arg \max}\limits_{\myVec{s} \in \mySet{S}^{\Nusers}} \Pdf{\myS | \myY} (\myVec{s} | \myVec{y}).
\eqspace
	\end{equation}
	
	The \ac{map} detector jointly recovers the symbols of all users  by searching over a set of $\CnstSize^{\Nusers}$ different possible input combinations, and thus becomes infeasible when the number of users $\Nusers$ grows. For example, when binary constellations are used, i.e., $\CnstSize = 2$, the number of different channel inputs is larger than $10^6$ for merely $\Nusers = 20$ users. 
	Furthermore, the \ac{map} decoder requires accurate knowledge of the channel model, i.e., the conditional distribution $\Pdf{\myY | \myS}$ must be fully known.  
	 A common strategy to implement joint decoding with affordable computational complexity, suitable for channels in which $\myY[i]$ is given by a linear transformation of $\myS[i]$ corrupted by additive noise, is interference cancellation \cite{andrews2005interference}.  Interference cancellation refers to a family of algorithms which implement joint decoding in an iterative fashion by recovering a subset of $\myS[i]$ based on the channel output as well as an estimate of the remaining interfering symbols. These algorithms facilitate the recovery of the subset of $\myS[i]$ from the channel output by canceling the effect of the estimated interference using knowledge of the channel parameters, and specifically, how each interfering symbol contributes to the channel output. In the sequel we present how interference cancellation, and specifically iterative \ac{sic} \cite{choi2000iterative}, can be learned from data without prior knowledge of the channel model $\Pdf{\myY |\myS}$.

\vspace{-0,2cm}
\subsection{Symbol Detector Derivation}
\label{subsec:DeepSICDerivation}
\vspace{-0,2cm}
	Here, we design a data-driven method for recovering  $\myS[i]$ from the channel output $\myY[i]$.  In particular, in our model the receiver  knows the constellation $\mySet{S}$, and that the channel is stationary and memoryless. We do not assume that the channel is linear nor  that the receiver knows the conditional probability measure $\Pdf{\myY | \myS}$.  
	Following the approach detailed in Section \ref{sec:ModelML}, we design our network to implement interference cancellation in a data-driven fashion. In particular, our proposed receiver is based on the iterative  \ac{sic} algorithm proposed in \cite{choi2000iterative}. 
	Therefore, as a preliminary step to designing the data-driven detector, we first review iterative \ac{sic}, after which we present DeepSIC, which is an \ac{ml}-based implementation of iterative \ac{sic}. 

	\subsubsection{Iterative Soft Interference Cancellation} 
\begin{figure}
		\centering
		\includegraphics[width = \columnwidth]{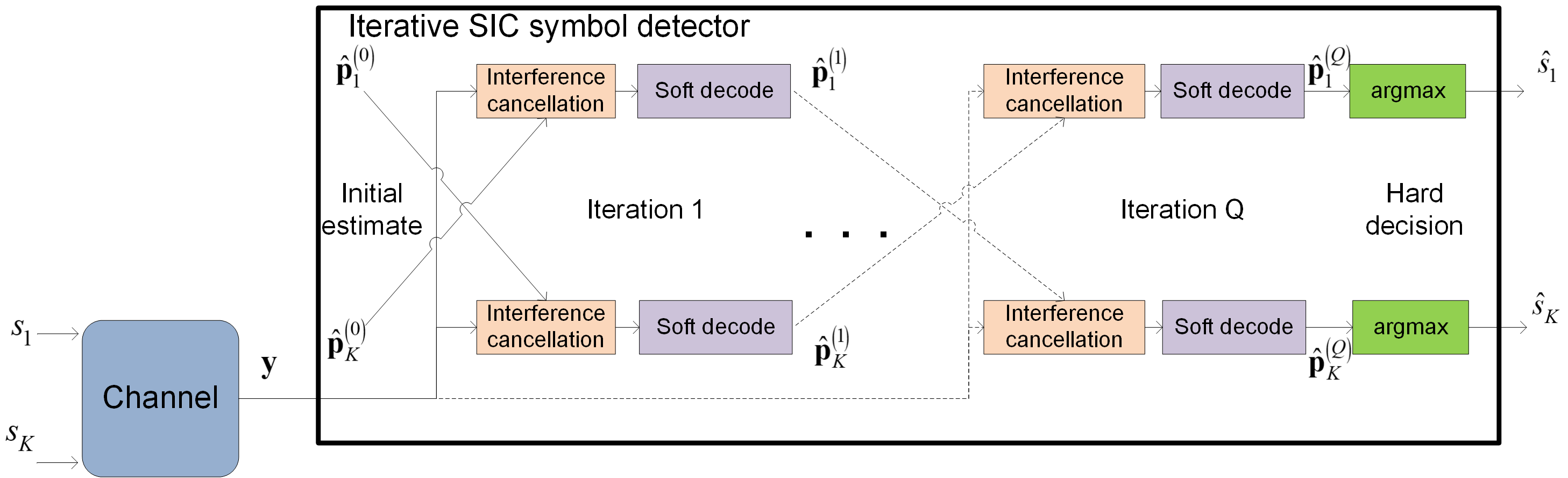}
		\vspace{-0.4cm}
		\caption{Soft iterative interference cancellation illustration.}
		\label{fig:SoftIC1}
	\end{figure}

	The iterative  \ac{sic} algorithm proposed in \cite{choi2000iterative} is a multiuser  detection method that combines multi-stage interference cancellation with soft decisions.  
	Broadly speaking, the detector operates in an iterative fashion where, in each iteration, an estimate of the conditional distribution of $S_k[i]$ given the observed $\myY[i] = \myVec{y}$ is generated for every user $k \in \NusersSet$ using the corresponding estimates of the interfering symbols $\{S_l[i]\}_{l \neq k}$ obtained in the previous iteration. Iteratively repeating this procedure refines the conditional distribution estimates, allowing the detector to accurately recover each symbol from the output of the last iteration.  This iterative procedure is illustrated in Fig.~\ref{fig:SoftIC1}.  
	
	To formulate the algorithm, we consider a channel whose output is obtained as a linear transformation of its input corrupted by \ac{awgn}, i.e., 
\eqspace
	\begin{equation}
	\label{eqn:Linear channel}
	\myY[i] = \myMat{H} \myS[i] + \myVec{W}[i],
\eqspace
	\end{equation}
	where $\myMat{H} \in \mySet{R}^{\Nantennas \times \Nusers}$ is an a-priori known channel matrix, and $\myVec{W}[i] \in \mySet{R}^{\Nusers}$ is a zero-mean  Gaussian vector with covariance $\SigW \myI_{\Nusers}$, independent of $\myS[i]$.
	
	Iterative \ac{sic} consists of $\Niter$ iterations. Each iteration indexed  $q \in \{1,2,\ldots, \Niter\} \triangleq \NiterSet$ generates $\Nusers$ distribution vectors $\hat{\myVec{p}}_k^{(q)} \in \mySet{R}^{\CnstSize}$, $k \in \NusersSet$. These vectors are computed from the channel output $\myVec{y}$ as well as the distribution vectors obtained at the previous iteration, $\{ \hat{\myVec{p}}_k^{(q-1)}\}_{k=1}^{\Nusers}$. The entries of  $\hat{\myVec{p}}_k^{(q)}$ are estimates of the distribution of $S_k[i]$ for each possible symbol in $\mathcal{S}$, given the channel output $\myY[i] = \myVec{y}$ and assuming that the interfering symbols $\{S_l[i]\}_{l \neq k}$ are distributed via $\{ \hat{\myVec{p}}_l^{(q-1)}\}_{l \neq k}$. 
	Every iteration consists of two steps, carried out in parallel for each user: {\em Interference cancellation}, and {\em soft decoding}. 
	Focusing on the $k$th user and the $q$th iteration, the interference cancellation stage first computes the expected values and variances of $\{S_l[i]\}_{l \neq k}$ based on  $\{ \hat{\myVec{p}}_l^{(q-1)}\}_{l \neq k}$. Letting $\{\alpha_j\}_{j=1}^{\CnstSize}$ be the indexed elements of the constellation set $\mySet{S}$, the expected values and variances are computed via
	$e_l^{(q-1)} = \sum_{\alpha_j \in \mySet{S}} \alpha_j \big( \hat{\myVec{p}}_l^{(q-1)}\big)_j$, 
	and 
	$v_l^{(q-1)} = \sum_{\alpha_j \in \mySet{S}}\big(  \alpha_j - e_l^{(q-1)}\big) ^2 \big( \hat{\myVec{p}}_l^{(q-1)}\big)_j$, 
	respectively. 
	The contribution of the interfering symbols from $\myVec{y}$ is then canceled by replacing them with $\{e_l^{(q-1)}\}$ and subtracting their resulting term. Letting $\myVec{h}_l$ be the $l$th column of $\myMat{H}$, the interference canceled channel output is given by
\eqspace
	\begin{align}
	\label{eqn:Cancelled1} 
	\myVec{Z}_k^{(q)}[i] 
	&= \myVec{Y}[i] \!- \!\sum\limits_{l \neq k} \myVec{h}_l e_l^{(q-1)}  
	= \myVec{h}_k S_k[i] \!+\! \sum\limits_{l \neq k} \myVec{h}_l (S_l[i] \!-\! e_l^{(q-1)}) \!+\! \myVec{W}[i].
\eqspace
	\end{align} 
	Substituting the channel output $\myVec{y}$ into \eqref{eqn:Cancelled1}, the realization of the interference canceled $\myVec{Z}_k^{(q)}[i]$, denoted  $\myVec{z}_k^{(q)}$, is obtained. 
	
	To implement soft decoding, it is assumed that $\tilde{W}_k^{(q)}[i] \triangleq \sum_{l \neq k} \myVec{h}_l (S_l[i]\!- \!e_l^{(q\!-\! 1)}) + \myVec{W}[i]$ obeys a zero-mean Gaussian distribution, independent of $S_k[i]$, and that its covariance is given by
	$\CovMat{\tilde{W}_k^{(q)}} = \SigW \myI_{\Nusers} + \sum_{l \neq k}v_l^{(q-1)}  \myVec{h}_l \myVec{h}_l^T$.
	Combining this assumption with \eqref{eqn:Cancelled1}, the conditional distribution of $\myVec{Z}_k^{(q)}$ given $S_k[i] = \alpha_j$ is multivariate Gaussian with mean value  $\myVec{h}_k \alpha_j$ and covariance $\CovMat{\tilde{W}_k^{(q)}}$. Since $\myVec{Z}_k^{(q)}[i] $ is given by a bijective transformation of $\myVec{Y}[i]$, it holds that $\Pdf{S_k | \myY}(\alpha_j | \myVec{y}) = \Pdf{S_k |  \myVec{Z}_k^{(q)}}(\alpha_j | \myVec{z}_k^{(q)})$ for each $\alpha_j \in \mySet{S}$ under the above assumptions. Consequently, the conditional distribution of $S_k[i]$ given $\myY[i]$ is approximated from the conditional distribution of $\myVec{Z}_k^{(q)}$ given $S_k[i]$ via Bayes theorem. Since the symbols are equiprobable, this estimated conditional distribution is computed as
\eqspace
	\begin{align}
	\left( \hat{\myVec{p}}_k^{(q)}\right)_j %
	&= \frac{\exp \left\{-\frac{1}{2} \left( \myVec{z}_k^{(q)} - \myVec{h}_k\alpha_j \right)^T\CovMat{\tilde{W}_k^{(q)}}^{-1} \left( \myVec{z}_k^{(q)} - \myVec{h}_k\alpha_j \right)   \right\} } { \sum\limits_{\alpha_{j'}\in\mySet{S} }\exp \left\{-\frac{1}{2} \left( \myVec{z}_k^{(q)} - \myVec{h}_k\alpha_{j'} \right)^T\CovMat{\tilde{W}_k^{(q)}}^{-1} \left( \myVec{z}_k^{(q)} - \myVec{h}_k\alpha_{j'} \right)   \right\} }.
	\label{eqn:CondDist2}
\eqspace
	\end{align}

	After the final iteration, the symbols are decoded by taking the symbol that maximizes the estimated conditional distribution for each user, i.e., 
\eqspace
	\begin{equation}
	\label{eqn:HardDet}
	\hat{s}_k = \mathop{\arg \max}\limits_{j \in \{1,\ldots,\CnstSize\}}\left( \hat{\myVec{p}}_k^{(\Niter)}\right)_j.
\eqspace
	\end{equation}
	The overall joint detection scheme is summarized below as Algorithm \ref{alg:Algo1SIC}.
	 The initial estimates $\{\hat{\myVec{p}}_k^{(0)}\}_{k=1}^{\Nusers}$ can be arbitrarily set. For example, these may be chosen based on a linear separate estimation of each symbol for $\myVec{y}$, as proposed in \cite{choi2000iterative}. 

\begin{algorithm}
	\caption{Iterative Soft Interference Cancellation Algorithm \cite{choi2000iterative}}
	\label{alg:Algo1SIC}
	\begin{algorithmic}[1]
		\STATE \underline{Input}: Channel output $\myVec{y}$.
		\STATE \underline{Initialization}: Set $q=1$, and generate an initial guess of  $\{\hat{\myVec{p}}_k^{(0)}\}_{k =1}^{\Nusers}$. 
		\STATE \label{stp:MF1a} Compute the expected values $\{e_k^{(q-1)}\}$ and variances $\{v_k^{q-1}\}$. 
		\STATE \label{stp:IC} {\em Interference cancellation:} For each $k \in \NusersSet$ compute $\myVec{z}_k^{(q)}$ via \eqref{eqn:Cancelled1}.
		\STATE \label{stp:SoftDec} {\em Soft decoding:} For each $k \in \NusersSet$, estimate   $\hat{\myVec{p}}_k^{(q)}$ via \eqref{eqn:CondDist2}.
		\STATE Set $q := q+1$. If $q \le \Niter $ go to Step \ref{stp:MF1a}.	
		
		\STATE  \underline{Output}: Hard decoded output $\hat{\myVec{s}}$, obtained via \eqref{eqn:HardDet}.
	\end{algorithmic}
\end{algorithm} 

Iterative \ac{sic} has several notable advantages as a joint decoding method: In terms of computational complexity, it replaces the joint exhaustive search over all different channel input combinations, required by the \ac{map} decoder \eqref{eqn:MAPMimo}, with a set of computations carried out separately for each user. Hence, its computational complexity only grows linearly with the number of users \cite{andrews2005interference}, making it feasible also with large values of $\Nusers$.  Unlike conventional separate decoding, in which the symbol of each user is recovered individually while treating the interference as noise, the iterative procedure refines the separate estimates sequentially, and the usage of soft values mitigates the effect of error propagation. Consequently, Algorithm \ref{alg:Algo1SIC} is capable of achieving performance approaching that of the \ac{map} detector, which is only feasible for small values of $\Nusers$.  

Iterative \ac{sic} is specifically designed for linear channels of the form \eqref{eqn:Linear channel}. In particular, the interference cancellation in Step \ref{stp:IC} of Algorithm \ref{alg:Algo1SIC} requires the contribution of the interfering symbols to be additive. This limits the application of the algorithm in non-linear channels. Additionally, the fact that the distribution of the interference canceled channel output $\myVec{Z}_k^{(q)}$ given $S_k[i]$ is approximated as Gaussian in Algorithm \ref{alg:Algo1SIC}  degrades the performance in channels which do not obey the linear Gaussian model \eqref{eqn:Linear channel}.
Furthermore, even when the channel obeys the linear model of \eqref{eqn:Linear channel}, iterative \ac{sic} requires full \ac{csi}, i.e.,  knowledge of the channel matrix $\myMat{H}$ and the noise variance $\SigW$, which may entail substantial overhead. 
The dependence  on accurate \ac{csi} and the assumption of linear channels are not unique to iterative \ac{sic}, and are in fact common to most interference cancellation based joint detection algorithms \cite{andrews2005interference}. These limitations motivate the design of a joint detector which exploits the computational feasibility of interference cancellation methods  while operating in a data-driven fashion. We specifically select iterative \ac{sic} since it is  capable of achieving \ac{map}-comparable performance, with a structure that can be readily converted to be data-driven. This is a result of the fact that its specific model-based computations, i.e., Steps \ref{stp:IC}-\ref{stp:SoftDec} in Algorithm \ref{alg:Algo1SIC}, can be naturally implemented using relatively simple \ac{ml} methods.  The resulting receiver, detailed in the following, integrates \ac{ml} methods into Algorithm \ref{alg:Algo1SIC}, allowing it to be implemented for arbitrary memoryless stationary channels without requiring a-priori knowledge of the channel model and its parameters. 

\subsubsection{Data-Driven Receiver Architecture}
\label{subsec:Derivation} 
Here, we present a receiver architecture that implements iterative \ac{sic} in a data-driven fashion. Following the approach of Section \ref{sec:ModelML}, we keep the overall structure of the iterative \ac{sic} algorithm, depicted in Fig. \ref{fig:SoftIC1}, while replacing the channel-model-based computations with dedicated suitable \acp{dnn}. To that aim, we note that  iterative \ac{sic}  can be viewed as a set of interconnected basic building blocks, each implementing the two stages of interference cancellation and soft decoding, i.e., Steps \ref{stp:IC}-\ref{stp:SoftDec} of Algorithm \ref{alg:Algo1SIC}. While the high level architecture of Fig. \ref{fig:SoftIC1} is ignorant of the underlying channel model, its basic building blocks are channel-model-dependent. In particular, interference cancellation requires the contribution of the interference to be additive, i.e., a linear model channel as in \eqref{eqn:Linear channel}, as well as full \ac{csi}, in order to cancel the contribution of the interference. Soft decoding requires complete knowledge of the channel input-output relationship in order to estimate the conditional probabilities via \eqref{eqn:CondDist2}. 

\begin{figure}
		\centering
		\includegraphics[width = \columnwidth]{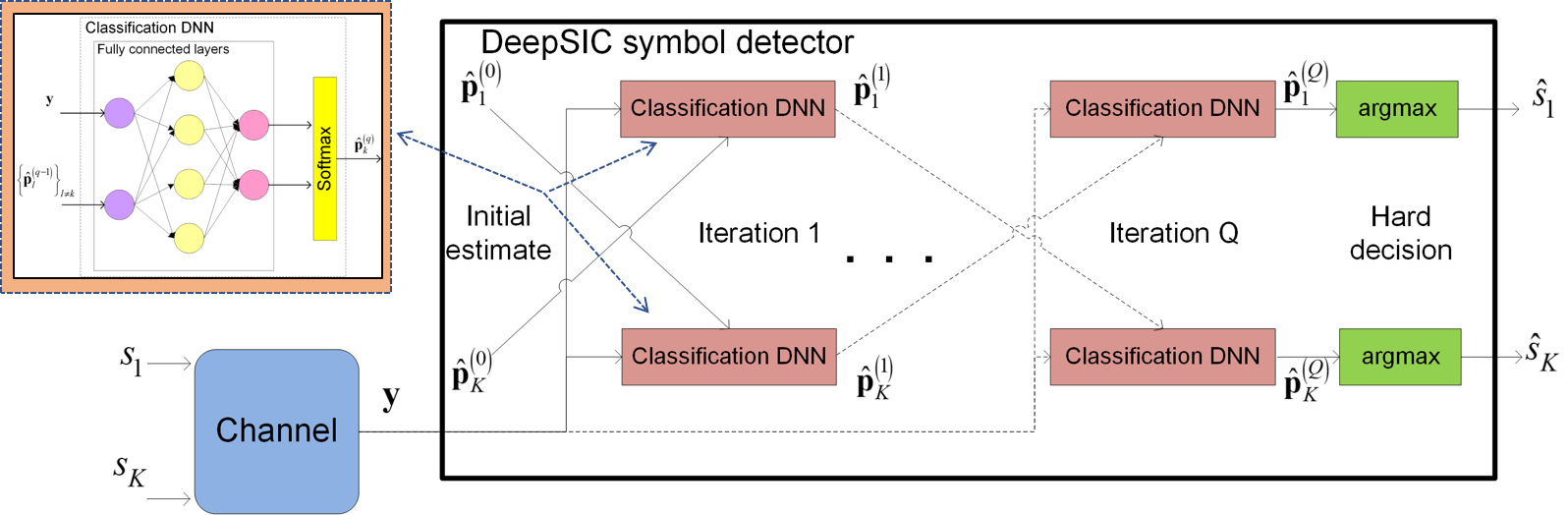} 
		\vspace{-0.4cm}
		\caption{Deep\ac{sic} illustration.}
		\label{fig:DeepSoftIC1}
	\end{figure}

Although each of these basic building blocks consists of two sequential procedures which are completely channel-model-based, we note that the purpose of these computations is to carry out a classification task. In particular, the $k$th building block of the $q$th iteration, $k \in \NusersSet$, $q  \in \NiterSet$, produces $\hat{\myVec{p}}_k^{(q)}$, which is an estimate of the conditional distribution of $S_k[i]$ given $\myY[i] = \myVec{y}$ based on $\{\hat{\myVec{p}}_l^{(q-1)}\}_{l\neq k}$. Such computations are naturally implemented by classification \acp{dnn}, e.g., fully-connected networks with a softmax output layer. 
Embedding these \ac{ml}-based conditional distribution computations into the iterative \ac{sic} block diagram in Fig. \ref{fig:SoftIC1} yields the overall receiver architecture depicted in Fig. \ref{fig:DeepSoftIC1}. We set the initial estimates $\{\hat{\myVec{p}}_k^{(0)}\}_{k=1}^{\Nusers}$ to represent a uniform distribution, i.e., $\big(\hat{\myVec{p}}_k^{(0)}\big)_j = \frac{1}{\CnstSize}$ for each $j \in \{1,2,\ldots,\CnstSize\}$ and $k \in \NusersSet$. The resulting data-driven implementation of Algorithm \ref{alg:Algo1SIC} is repeated below as Algorithm \ref{alg:AlgoDeepSIC}. Note that the model-based Steps \ref{stp:MF1a}-\ref{stp:SoftDec} of Algorithm \ref{alg:Algo1SIC}, which estimate the conditional distributions, are replaced with the \ac{ml}-based conditional distribution estimation Step \ref{stp:DeepSoftDec} in Algorithm \ref{alg:AlgoDeepSIC}.


\begin{algorithm}
	\caption{Deep Soft Interference Cancellation (Deep\ac{sic})}
	\label{alg:AlgoDeepSIC}
	\begin{algorithmic}[1]
		\STATE \underline{Input}: Channel output $\myVec{y}$.
		\STATE \underline{Initialization}: Set $q=1$, and generate an initial guess of the conditional distributions $\{\hat{\myVec{p}}_k^{(0)}\}_{k =1}^{\Nusers}$.  
		\STATE \label{stp:DeepSoftDec} {\em \ac{ml}-based conditional distribution estimation:} For each $k \in \NusersSet$, estimate the conditional distribution $\hat{\myVec{p}}_k^{(q)}$ from  $\myVec{y}$ and $\{\hat{\myVec{p}}_k^{(q-1)}\}_{l\neq k}$ using the $(q,k)$th classification \ac{dnn}.
		\STATE Set $q := q+1$. If $q \le \Niter $ go to Step \ref{stp:DeepSoftDec}.	
		
		\STATE  \underline{Output}: Hard decoded output $\hat{\myVec{s}}$, obtained via \eqref{eqn:HardDet}.
	\end{algorithmic}
\end{algorithm} 

A major advantage of using classification \acp{dnn} as the basic building blocks in Fig. \ref{fig:DeepSoftIC1} stems from the fact that such \ac{ml}-based methods are capable of accurately computing conditional distributions in   complex non-linear setups without requiring a-priori knowledge of the channel model and its parameters. Consequently, when these building blocks are trained to properly implement their classification task, the  receiver essentially realizes iterative soft interference cancellation for arbitrary channel models in a data-driven fashion. 

\subsubsection{Training the \acp{dnn}}
\label{subsec:Training}
\vspace{-0.1cm}	
In order for the \ac{ml}-based receiver structure of Fig.~\ref{fig:DeepSoftIC1} to  reliably implement joint decoding, its building block classification \acp{dnn} must be properly trained. Here, we consider two possible approaches to train the receiver based on a set of $\Ntraining$ pairs of channel inputs and their corresponding outputs, denoted $\{\myVec{s}_i, \myVec{y}_i \}_{i=1}^{\Ntraining}$: {\em End-to-end training}, and {\em sequential training}.

{\bf End-to-end training}: The first approach jointly trains the entire network, i.e., all the building block \acp{dnn}. Since the output of the deep network is the set of conditional distributions  $\{\hat{\myVec{p}}_k^{(\Niter)}\}_{k=1}^{\Nusers}$, where each  $\hat{\myVec{p}}_k^{(\Niter)}$ is used to estimate $S_k[i]$, we use the sum cross entropy  as the training objective. Let 
$\myVec{\theta}$ be the network parameters, and 
$\hat{\myVec{p}}_k^{(\Niter)}(\myVec{y}, \alpha; \myVec{\theta} )$ be the entry of $\hat{\myVec{p}}_k^{(\Niter)}$ corresponding to $S_k[i] = \alpha$ when the input to the network parameterizd by $\myVec{\theta}$ is $\myVec{y}$.
The sum cross entropy loss is
\eqspace
\begin{equation}
\label{eqn:SumCE}
\mathcal{L}_{\rm SumCE} (\myVec{\theta})= \frac{1}{\Ntraining}\sum_{i=1}^{\Ntraining}
\sum_{k=1}^{\Nusers} -\log \hat{\myVec{p}}_k^{(\Niter)}\big(\myVec{y}_i, (\myVec{s}_i)_k ; \myVec{\theta}\big).
\eqspace
\end{equation} 

Training the receiver in Fig. \ref{fig:DeepSoftIC1} in an end-to-end manner based on the loss \eqref{eqn:SumCE} jointly updates the coefficients of all the $\Nusers \cdot \Niter$ building block \acp{dnn}. Since for a large number of users, training so many parameters simultaneously is expected to require a large labeled set, we next propose a sequential training approach. 

{\bf Sequential training}: To allow the network to be trained with a reduced number of training samples, we note that the goal of each building block \ac{dnn} does not depend on the iteration index: The $k$th building block of the $q$th iteration outputs a soft estimate of $S_k[i]$ for each $q \in \NiterSet$. Therefore, each building block  can be trained individually to minimize the cross entropy loss. To formulate this objective, let  
$\myVec{\theta}_{k}^{(q)}$ be the parameters of the $k$th \ac{dnn} at iteration $q$, and write 
$\hat{\myVec{p}}_k^{(q)}\big(\myVec{y}, \{\hat{\myVec{p}}_l^{(q-1)}\}_{l\neq kl}, \alpha; \myVec{\theta}_{k}^{(q)}\big)$ as the entry of $\hat{\myVec{p}}_k^{(q)}$ corresponding to $S_k[i] = \alpha$ when  its inputs are $\myVec{y}$ and  $\{\hat{\myVec{p}}_l^{(q-1)}\}_{l\neq l}$. The cross entropy loss is
\eqspace
 \begin{equation}
 \label{eqn:CE}
 \mathcal{L}_{\rm CE}(\myVec{\theta}_{k}^{(q)})  = \frac{1}{\Ntraining}\sum_{i=1}^{\Ntraining}
  -\log \hat{\myVec{p}}_k^{(q)}\big(\myVec{y}_i, \{\hat{\myVec{p}}_{i,l}^{(q-1)}\}_{l\neq k}, (\myVec{s}_i)_k ; \myVec{\theta}_{k}^{(q)}\big),
\eqspace
 \end{equation}	
 where $\{\hat{\myVec{p}}_{i,l}^{(q-1)}\}$ represent the estimated probabilities associated with $\myVec{y}_i$ computed at the previous iteration.
 The problem with training each \ac{dnn} individually is that the soft estimates $\{\hat{\myVec{p}}_{i,l}^{(q-1)}\}$ are not provided as part of the training set. This challenge can be tackled by training the \acp{dnn} corresponding to each layer in a sequential manner, where for each layer the outputs of the trained \ac{dnn} corresponding to the previous iterations are used as the soft estimates fed as training samples. 
 
Sequential training uses the $\Ntraining$ input-output pairs to train each \ac{dnn} individually. Compared to the end-to-end training that utilizes the training samples to learn the complete set of parameters, which can be quite large, sequential training uses the same data set  to learn a significantly smaller number of parameters, reduced by a factor of $\Nusers\cdot \Niter$, multiple times. Consequently, this approach is expected to require much fewer training samples, at the cost of a longer learning procedure for a given training set, due to its sequential operation, and possible performance degradation as the building blocks are not jointly trained. This behavior is numerically demonstrated in the simulation study detailed in Section \ref{subsec:DeepSICSims}. 


\vspace{-0,2cm}
\subsection{Numerical Evaluations}
\label{subsec:DeepSICSims}
\vspace{-0,2cm}
	In the following section we numerically evaluate Deep\ac{sic}. We train  Deep\ac{sic}  using the ADAM optimizer \cite{kingma2014adam} with a relatively small training set of $5000$ training samples, and tested over $20000$ symbols. In the implementation of the \ac{dnn}-based building blocks of Deep\ac{sic},  we used a different fully-connected network for each training method: For end-to-end training, where all the building blocks are jointly trained, we used a compact network consisting of a  $(\Nantennas + \Nusers - 1) \times 60$ layer followed by ReLU activation and a $60 \times \CnstSize$  layer. 
	For sequential training, which sequentially adapts subsets of the building blocks and can thus tune a more parameters using the same training set  (or, alternatively,  requires a smaller training set) compared to end-to-end training, we used three fully-connected layers:  An $(\Nantennas + \Nusers - 1) \times 100$ first layer, a $100 \times 50$ second layer, and a $50 \times \CnstSize$ third layer, with a sigmoid and a ReLU intermediate activation functions, respectively.  
	In both iterative \ac{sic} as well as Deep\ac{sic}, we set the number of iterations to $\Niter = 5$.
	
		We first consider a linear \ac{awgn} channel as in \eqref{eqn:Linear channel}. Recall that iterative \ac{sic} as well as previously proposed unfolding-based data-driven \ac{mimo} receivers \cite[Sec. II]{balatsoukas2019deep}, are all designed for such channels. Consequently,  the following study compares Deep\ac{sic} in terms of performance and robustness to competing detectors in a scenario for which these previous schemes are applicable. 
	In particular, we evaluate the \ac{ser} of the following \ac{mimo} detectors:
    The \ac{map} detector, given by \eqref{eqn:MAPMimo}; The iterative \ac{sic} algorithm (Algorithm \ref{alg:Algo1SIC}); Deep\ac{sic} with the sequential training method, referred to in the following as {\em Seq. Deep\ac{sic}};  Deep\ac{sic} with end-to-end training based on the sum cross entropy loss \eqref{eqn:SumCE}, referred to henceforth as {\em E2E Deep\ac{sic}}; and the unfolding-based {\em DetNet} \ac{mimo} detector  \cite{samuel2019learning}.
	
	The model-based \ac{map} and iterative \ac{sic} detectors, as well as DetNet \cite{samuel2019learning}, all require \ac{csi}, and specifically, accurate knowledge of the channel matrix $\myMat{H}$.  Deep\ac{sic} operates without a-priori knowledge of the channel model and its parameters, learning the decoding mapping from a training set sampled from the considered input-output relationship. In order to compare the robustness of the  detectors to \ac{csi} uncertainty, we also evaluate them when the receiver has access to an estimate of $\myMat{H}$ with entries corrupted by i.i.d. additive Gaussian noise whose variance is given by $\SigE$ times the magnitude of the corresponding entry, where $\SigE > 0$ is referred to as the {\em error variance}. For Deep\ac{sic}, which is model-invariant, we compute the \ac{ser} under \ac{csi} uncertainty by using a training set whose samples are randomized from a channel in which the true $\myMat{H}$ is replaced with its noisy version. 
	
\begin{figure}
	\centering
	\begin{minipage}{0.45\textwidth}
		\centering
		\scalebox{0.38}{\includegraphics{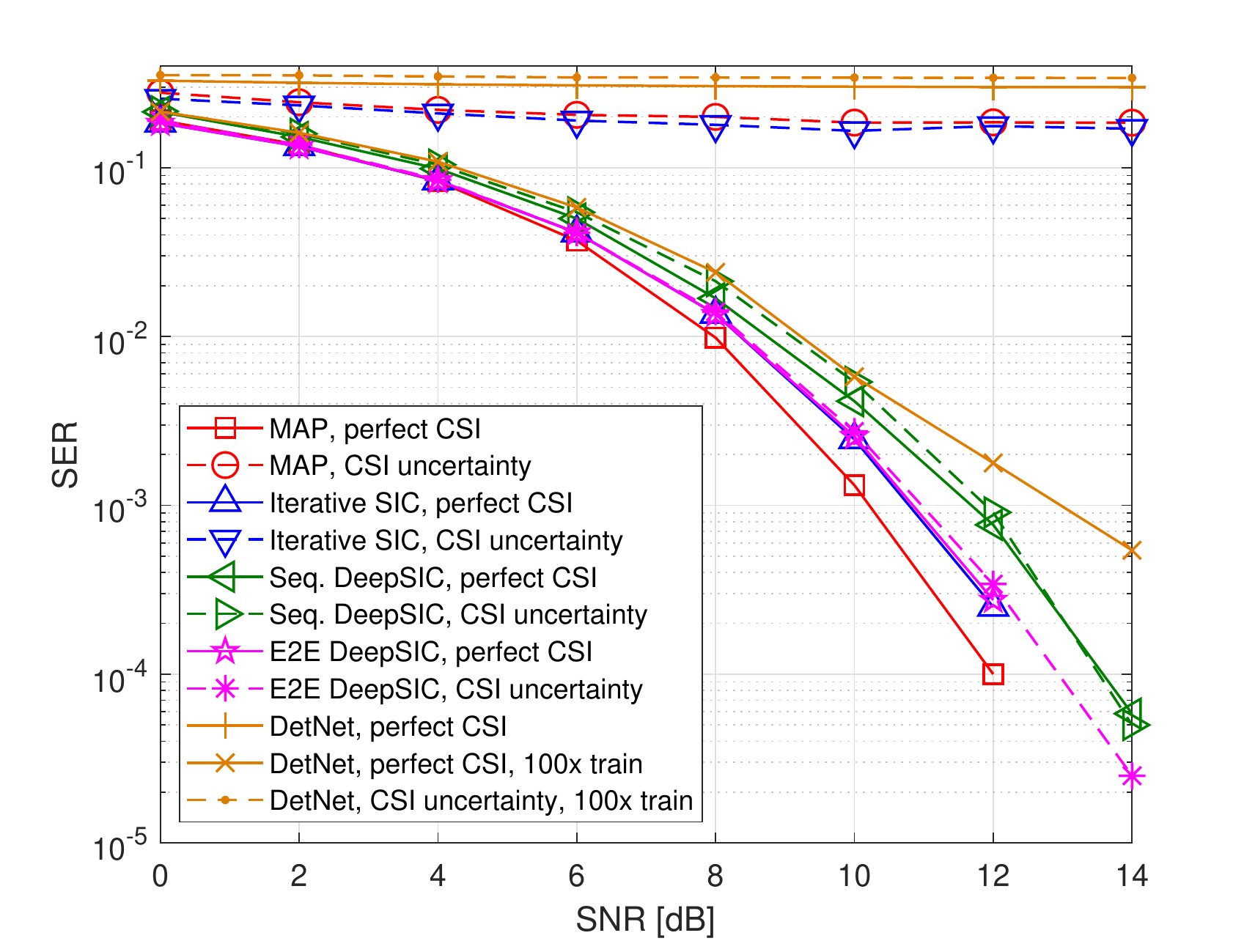}}
		\figSpace
		\caption{ $6\times6$ \ac{awgn} channel.
		}
		\label{fig:AWGN6} 	
	\end{minipage}
	$\quad$
	\begin{minipage}{0.45\textwidth}
		\centering
		\scalebox{0.38}{\includegraphics{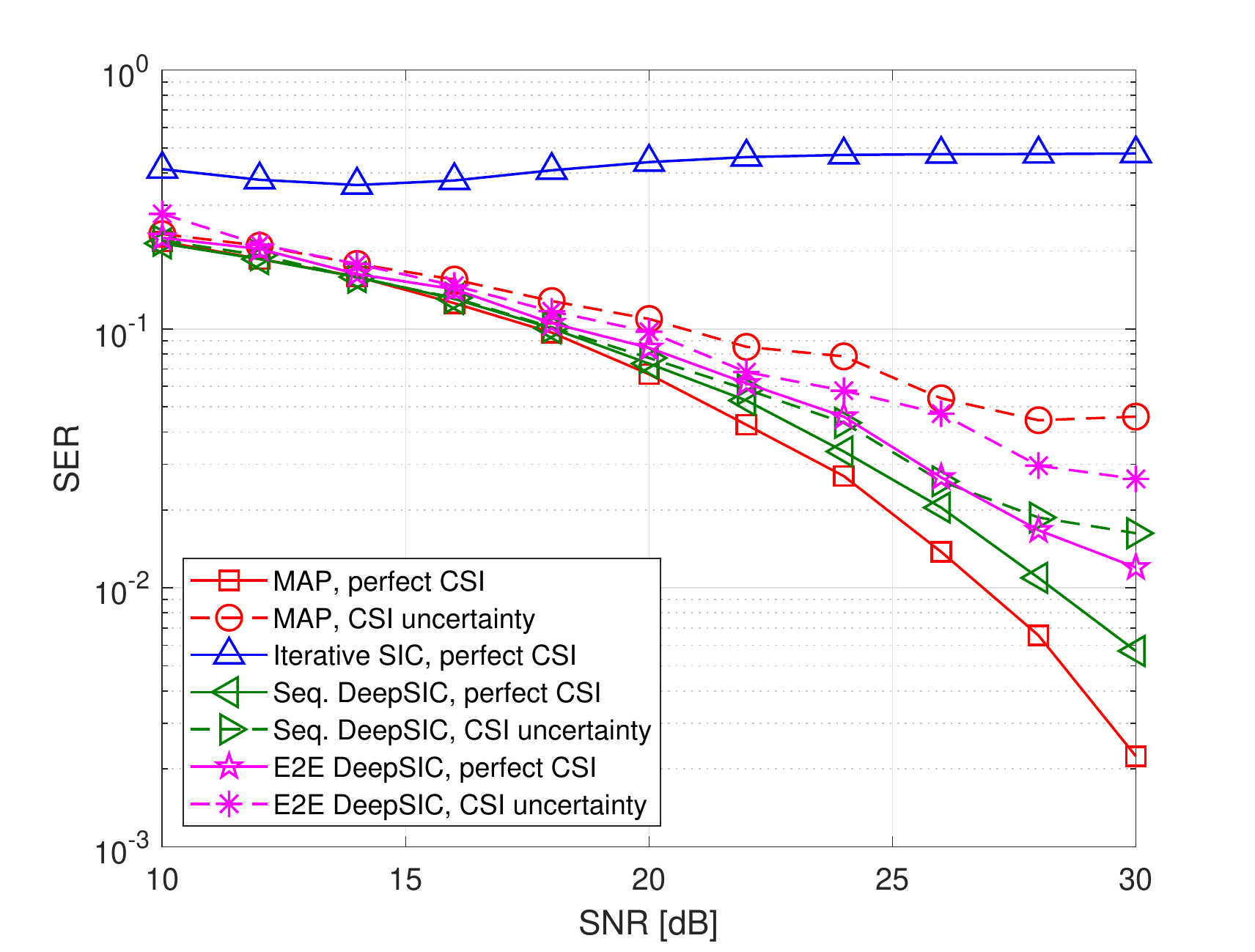}}
    	\figSpace
    	\caption{$4\times 4$ Poisson channel.
    	}
    	\label{fig:Poisson4} 
	\end{minipage}
\end{figure}

	We simulate $6 \times 6$ linear Gaussian channel, i.e., $\Nusers = 6$ users and $\Nantennas = 6$ receive antennas. The symbols are randomized from a \ac{bpsk} constellation, namely, $\mySet{S} = \{-1, 1\}$ and $\CnstSize = |\mySet{S}| = 2$. The channel matrix $\myMat{H}$ models spatial exponential decay, and its entries are given by
	$\left( \myMat{H}\right)_{i,k} = e^{-|i-j|}$, for each $i \in \{1,\ldots, \Nantennas\}$,$ k \in \NusersSet$.  
	For each channel, the \ac{ser} of the considered receivers is evaluated for both perfect \ac{csi}, i.e., $\SigE = 0$, as well as \ac{csi} uncertainty, for which we use $\SigE = 0.1$.  The  evaluated \ac{ser}  versus the \ac{snr}, defined as $1/\SigW$, is depicted in Fig. \ref{fig:AWGN6}. 
	
	Observing Fig. \ref{fig:AWGN6}, we note that the performance of Deep\ac{sic} with end-to-end training approaches that of the model-based iterative \ac{sic} algorithm, which is within a small gap of the optimal \ac{map} performance. This demonstrates the ability of Deep\ac{sic} to implement iterative \ac{sic} in a data-driven fashion. The sequential training method, whose purpose is to allow Deep\ac{sic} to train with smaller data sets compared to end-to-end training, also achieves \ac{ser} which is comparable to iterative \ac{sic}. In the presence of \ac{csi} uncertainty, Deep\ac{sic} is observed to substantially outperform the model-based iterative \ac{sic} and \ac{map} receivers, as well as DetNet operating with a noisy version of $\myMat{H}$ and trained with a hundred times more samples. In particular, it follows from Fig. \ref{fig:AWGN6} that a relatively minor  error of variance   $\SigE = 0.1$ severely deteriorates the performance of the model-based methods, while   Deep\ac{sic} is hardly affected by the same level of \ac{csi} uncertainty. 

		Next, we consider a Poisson channel. We use $\Nusers = 4$ and $\Nantennas = 4$. Here, the symbols are randomized from an on-off keying for which $\mySet{S} = \{0,1\}$. The entries of the channel output are related to the input via the conditional distribution
\eqspace
	\begin{equation}
	\label{eqn:PoissonChannel}
	\left( \myY[i]\right)_j  | \myS[i]  \sim \mathbb{P}\left( \frac{1}{\sqrt{\SigW}}\left( \myMat{H} \myS[i]\right)_j  + 1\right), \qquad j \in \{1,\ldots, \Nantennas\},
\eqspace
	\end{equation}
	where $\mathbb{P}(\lambda)$ is the Poisson distribution with parameter $\lambda > 0$. 
	
	The achievable \ac{ser} of Deep\ac{sic} versus \ac{snr} under both perfect \ac{csi} as well as \ac{csi} uncertainty with error variance $\SigE = 0.1$ is compared to the \ac{map} and iterative \ac{sic} detectors in Fig. \ref{fig:Poisson4}. Observing Fig. \ref{fig:Poisson4}, we again note that the performance of Deep\ac{sic} is only within a small gap of the \ac{map} performance with perfect \ac{csi}, and that the data-driven receiver is more robust to \ac{csi} uncertainty compared to the model-based \ac{map}. 
	In particular, Deep\ac{sic} with sequential training, which utilizes a deeper network architecture for each building block, outperforms here end-to-end training with basic two-layer structures for the conditional distribution estimation components. We conclude that under such non-Gaussian channels, more complex \ac{dnn} models are required to learn to cancel interference and carry out soft detection accurately. This further emphasizes the gain of our proposed sequential approach for training each building block separately, thus allowing to train an overall deep architecture using a limited training set based on the understanding of the role of each of its components. 
	 Furthermore,  iterative \ac{sic}, which is designed for linear Gaussian channels \eqref{eqn:Linear channel} where interference is additive, achieves very poor performance when the channel model is substantially different from \eqref{eqn:Linear channel}. 

	Our numerical results demonstrate the ability of Deep\ac{sic} to achieve excellent performance through learning from data for statistical models where model-based interference cancellation is effectively inapplicable.

\vspace{-0,2cm}
\section{Discussion and Conclusion}
\label{sec:Conclusion}
\vspace{-0,2cm}
	We reviewed an \ac{ml}-based approach for designing symbol detection algorithms for communication systems. This approach introduces \ac{ml} into well-known algorithms such as Viterbi, BCJR, and \ac{mimo} \ac{sic}, by identifying the computations that require full knowledge of the underlying channel input-output statistical relationships, and replacing them with \ac{ml}-based algorithms. The resulting architecture combines \ac{ml}-based processing with conventional symbol detection schemes. Our numerical results demonstrate that the performance of data-driven ViterbiNet, BCJRNet, and DeepSIC approaches the optimal performance of the equivalent \ac{csi}-based versions, and outperform previously proposed \ac{ml}-based symbol detectors using a small amount of training data. It is also illustrated that these algorithms are capable of operating in the presence of \ac{csi} uncertainty with little performance degradation. Since our approach relies on small neural networks that can be trained quickly to achieve close to optimal performance with only a few thousand training symbols, it paves the way to the possibility of online training using pilot sequences. 

As part of future work, we will be extending our approach to general factor graph methods as well as using meta learning to improve the adaptability of our algorithms to changing channels. Finally, we will also investigate how different density estimation methods can be used to further improve the training process.

\vspace{-0,2cm}
\bibliographystyle{IEEEtran}
\bibliography{IEEEabrv,refs}

\end{document}